\documentclass[11pt,a4paper]{article}
\usepackage{jheppub}
\usepackage{ifpdf}
\usepackage[usenames,dvipsnames]{xcolor}
\usepackage{ae}
\usepackage[T1]{fontenc}
\usepackage[ansinew]{inputenc}
\usepackage{mathrsfs}
\usepackage{amsmath}
\usepackage{amssymb}
\usepackage{graphicx}
\usepackage{color}
\definecolor{darkblue}{cmyk}{0.9,0.9,0,0}
\definecolor{darkgreen}{rgb}{0,0.55,0}

\usepackage{epsfig}
\usepackage{amsmath}

\newcommand{\comment}[1]{}

\newcommand{\beq}{\begin{equation}}
\newcommand{\eeq}{\end{equation}}
\newcommand{\beqq}{\begin{equation*}}
\newcommand{\eeqq}{\end{equation*}}
\newcommand\beqa{\begin{eqnarray}}
\newcommand\eeqa{\end{eqnarray}}
\newcommand\beqaa{\begin{eqnarray*}}
	\newcommand\eeqaa{\end{eqnarray*}}
\newcommand\bea{\begin{array}}
	\newcommand\eea{\end{array}}

\def\XXint#1#2#3{{\setbox0=\hbox{$#1{#2#3}{\int}$ }
		\vcenter{\hbox{$#2#3$ }}\kern-.5\wd0}}

\def\XXint#1#2#3{{\setbox0=\hbox{$#1{#2#3}{\int}$}
		\vcenter{\hbox{$#2#3$}}\kern-.5\wd0}}

\newcommand{\nn}{\nonumber}
\newcommand{\com}[1]{(*{\textbf{#1}}*)}

\newcommand{\neqa}{\nonumber\end{eqnarray}}
\newcommand{\la}[1]{\label{#1}}

\newcommand{\eq}[1]{(\ref{#1})}

\def\tr{{\rm tr~}}

\newcommand{\hs}{\frac{\sqrt{3}}{2}}
\renewcommand{\d}{\partial}

\newcommand{\<}{{\langle}}
\renewcommand{\>}{{\rangle}}

\newcommand{\cB}{{\cal B}}

\newcommand{\cL}{{\cal L}}

\newcommand{\re}{\relax{\rm I\kern-.18em R}}

\renewcommand{\sp}{p\hspace{-.40em}/}

\def\su2{{SU(2)}}

\def\[{\left[}
\def\]{\right]}

\def\s{\sigma}

\def\({\left(}
\def\){\right)}
\def\[{\left[}
\def\]{\right]}

\def\<{\langle}
\def\>{\rangle}

\def\cB{{\cal B}}
\def\bC{{\bf C}}
\def\cO{{\cal O}}

\def\cX{{\cal X}}

\def\s*{\ *_{\!\!\!\!\!\!\!\!\!\,_{\,_\text{\scriptsize{sym}}}}}
\def\hs*{\ \hat{*}_{\!\!\!\!\!\!\!\!\!\,_{\,_\text{\scriptsize{sym}}}}}
\def\d{\partial}

\def\i2{\frac{i}{2}}

\def\bq{{\bf q}}

\def\spi{\relax{\rm \pi\kern-0.5em /}}
\def\sA{\relax{\rm A\kern-0.5em /}}
\def\sp{\relax{\rm p\kern-0.5em /}}
\def\sd{\relax{\rm \d\kern-0.5em /}}
\def\sk{\relax{\rm k\kern-0.5em /}}
\def\sn{\relax{\rm n\kern-0.5em /}}
\def\sl{\relax{\rm l\kern-0.5em /}}
\def\sP{\relax{\rm P\kern-0.7em /}}
\def\sBethe{\relax{\rm \Bethe\kern-0.5em /}}
\def\cN{{\cal N}}

\def\bC{{\bold C}}

\numberwithin{equation}{section}

\usepackage{varioref}
\usepackage{makeidx}
\makeindex

\title{Quantum Fishchain in $AdS_5$}
\author[a,b]{Nikolay Gromov}
\author[c,d]{Amit Sever}%
\affiliation[a]{%
Mathematics Department, King's College London,
The Strand, London WC2R 2LS, UK
}%
\affiliation[b]{St.Petersburg INP, Gatchina, 188 300, St.Petersburg,
Russia}
\affiliation[c]{
School of Physics and Astronomy, Tel Aviv University, Ramat Aviv 69978, Israel}
\affiliation[d]{
CERN, Theoretical Physics Department, 1211 Geneva 23, Switzerland
}

\emailAdd{nikolay.gromov@kcl.ac.uk}
\emailAdd{amit.sever@cern.ch}
\abstract{In our previous paper we derived the holographic dual of the planar fishnet CFT in four dimensions. 
The dual model becomes classical in the strongly coupled regime of the CFT and takes the form of an integrable chain of particles in five dimensions. 
Here we study the theory at the quantum level. By applying the canonical quantization procedure with constraints, we show that the model describes a quantum chain of particles propagating in $AdS_5$. We prove the duality at the full quantum level in the ${\mathfrak u}(1)$ sector and reproduce exactly the spectrum for the cases when it is known analytically.
}
\begin{document}
\begin{flushright}
CERN-TH-2019-143
\end{flushright}    

\maketitle
\newpage
\section{Introduction}

After two decades of intense study, we have achieved a good understanding of the holographic principle. 
Nevertheless, there are still a number of important aspects of it that beg for a better understanding, such as the holographic map and the corresponding emergence of bulk locality. Finding the holographic dual is always a guesswork which requires a number of elaborate tests. There are only few firmly-established examples of such a map which are all heavily based on supersymmetry. 

One possible strategy to understand the general holographic map is to derive such explicit duality in a specific theory and then deduce from it general principles. 
For that one should find a non-trivial interacting analytically tractable model which would capture the general structure. The planar fishnet~\cite{Gurdogan:2015csr,Zamolodchikov:1980mb} is an example of such a model. 
It is a four-dimensional interacting CFT~\cite{Grabner:2017pgm}, with a free parameter $\xi$ that controls the strength of the interaction. The standard perturbation theory has a finite radius of convergence in $\xi$ and the corresponding Feynman perturbative expansion is relatively simple and can be computed exactly to all orders in some cases~\cite{Gromov:2017cja,Grabner:2017pgm,Gromov:2018hut}. 
Importantly, the fishnet model is continuously connected to ${\cal N}=4$ SYM theory. The connection between the two models can be realised as a double scaling limit of 
$\gamma$-deformation of ${\cal N}=4$ SYM~\cite{Gurdogan:2015csr}.
Another alternative way to obtain the fishnet is by taking a limit in the parameter space of some observables, such as Wilson lines with cusps in ${\cal N}=4$ SYM itself~\cite{Erickson:1999qv,Correa:2012nk}. Hence, a derivation of the holographic dual for the fishnet should set the ground for a derivation of the dual also for the parent ${\cal N}=4$ SYM theory.

In our previous paper~\cite{Gromov:2019aku} we have derived from the first principles the strong coupling dual of the fishnet CFT, which we call {\it the fishchain model}. In the ${\mathfrak{u}}(1)$ sector it is given by a chain of classical point-like particles that propagate on the lightcone of ${\mathbb R}^{1,5}$ with the nearest-neighbour interaction. Notably, this model is also classically integrable. 
In this paper we show that the quantization of this classical model produces the full quantum duality, valid at a finite value of the `t Hooft coupling $\xi$. As opposed to the classical fishchain, we find that the quantum fishchain lives in $AdS_5$. For simplicity, in this paper we work in the ${\mathfrak u}(1)$ sector of all operators, which we describe in the next section. The generalisation of our model beyond this sector will be discussed in \cite{magnons}.

The paper is organised as follows: In section \ref{sec2} we review the fishnet model and its strong coupling fishchain dual. In section \ref{sec3} we canonically quantize the fishchain model. In section \ref{sec4} we test the quantum duality by reproducing the finite coupling spectrum of the fishnet model for operators of length one and two. In section \ref{sec5} we present an alternative derivation of the duality and prove it for any length of the chain. Finally, in section \ref{sec6} we prove the quantum integrability of the fishchain model. We conclude in section \ref{sec7} with a short discussion.

\section{Review of the fishnet model and its fishchain dual}\la{sec2}

To set the ground for the paper and to fix our notation, we now review the relevant aspects of the fishnet model and its dual fishchain model. In particular, we present the fishchain model in the form that is most convenient for its quantization. We also introduce the ${\mathfrak{u}}(1)$ sub-sector, to which we restrict ourselves in this paper.

\subsection{The fishnet model}
The {\it fishnet model} \cite{Gurdogan:2015csr} is a QFT in four-dimensional spacetime. It consists of two complex $N\times N$ matrix scalar fields, $\phi_1$ and $\phi_2$. Its dynamics is dictated by the action\footnote{Here we have suppressed double trace interactions which are not relevant non-perturbatively \cite{Fokken:2013aea,Gromov:2018hut}.}
\beq\la{fishnet}
\cL_{4D} = N\,{\rm tr}\(|\d\phi_1|^2+|\d\phi_2|^2
+ (4\pi)^2 \xi^2 \phi_1^\dagger\phi_2^\dagger\phi_1\phi_2\)\,.
\eeq
The two scalars transform in the adjoint of $SU(N)$ and are charged under $U(1)_1$, $U(1)_2$ correspondingly.
In this paper we consider
$N$ to be taken large, while $\xi^2$ is a fixed arbitrary complex number. We refer to $\xi^2$ as the fishnet 't Hooft coupling.\footnote{Not to be confused with the 't Hooft coupling of the parent ${\cal N}=4$ SYM theory, $\lambda$.} In this limit, the model was shown to be conformal and integrable \cite{Fokken:2013aea,Gurdogan:2015csr,Gromov:2017cja,Grabner:2017pgm}. 
Since the interaction term in the action is not real, the theory described by (\ref{fishnet}) is not unitary.
Non-unitary CFTs appear in a condensed matter context~(see for example \cite{Couvreur:2016mbr} and references therein).
There are very few methods to study non-unitary CFTs in general. The most actively developing bootstrap methods are usually not applicable for such models. Nevertheless, integrability and holography can be used rather efficiently to study \eq{fishnet}.

In this paper we consider a subset of all states in what is known as the ${\mathfrak u}(1)$ sector of the model. It consists of operators with zero $U(1)_2$ charge. Such operators at weak coupling are of the form
\beq\la{U1}
\cO_J=\tr[\d^m\phi_1^J(\phi_2\phi_2^\dagger)^n\dots]+\dots
\eeq
containing any number of derivatives, $J$-scalar fields $\phi_1$ and any {\it neutral} combination of $\phi_2$ and $\phi_2^\dagger$.

For such operators we consider the following $(J+1)$--point correlator
\beq\la{wf}
\varphi_{\cO_J}(x_0;x_1,...,x_J)\equiv\<\cO_J(x_0){\rm tr}[\phi_1^\dagger(\vec x_1)\dots\phi_1^\dagger(\vec x_J)]\>\;.
\eeq
This correlator will play an important role in what follows.
The knowledge of this correlator is equivalent to the knowledge of the initial operator. Any other planar correlators, such as those with $\phi_2\phi_2^\dagger$'s, can be expressed in terms of this one.  Hence, we refer to it as ``{\it the CFT wave-function}". The corresponding off-shell ``{\it CFT norm}" between any two such CFT wave-functions in the ${\mathfrak u}(1)$ sector is given by\footnote{When taking the conjugate of the wave function in (\ref{CFTnorm}), one does not conjugate the 't Hooft coupling $\xi^2$. This relation is a straight forward generalization of the $J=2$ case considered in \cite{Grabner:2017pgm} to any length.}
\beq\la{CFTnorm}
\<\!\!\<\varphi_{\widetilde\cO}|\varphi_\cO\>\!\!\>\equiv\int\prod_{i=1}^J{d^4x_i\over4\pi^2}\,\bar\varphi_{\widetilde\cO}(x_0';x_1,\dots,x_J)\prod_{j=1}^J\(-
\Box_{j}\)\varphi_\cO(x_0;x_1,\dots,x_J)
\eeq
In particular, the graph building operator is self-adjoint with respect to this norm. Given the wave function $\varphi_\cO$ we can compute any correlation function of the operator $\cO$. For example, the two point function is given by
\beq
\<\widetilde\cO_J\,\cO_J\>={\<\!\!\<\varphi_{\widetilde\cO_J}|\varphi_{\cO_J}\>\!\!\>\over4J\log(\epsilon)\,\xi\d_\xi\Delta_{\cO_J}}\ ,
\eeq
where $\epsilon$ is a small UV cutoff length scale. In section \ref{sec5} we will also show that the  correlator \eq{wf} is indeed directly related to the wave function of the dual model.

\begin{figure}
	\centering
	\includegraphics[scale=1]{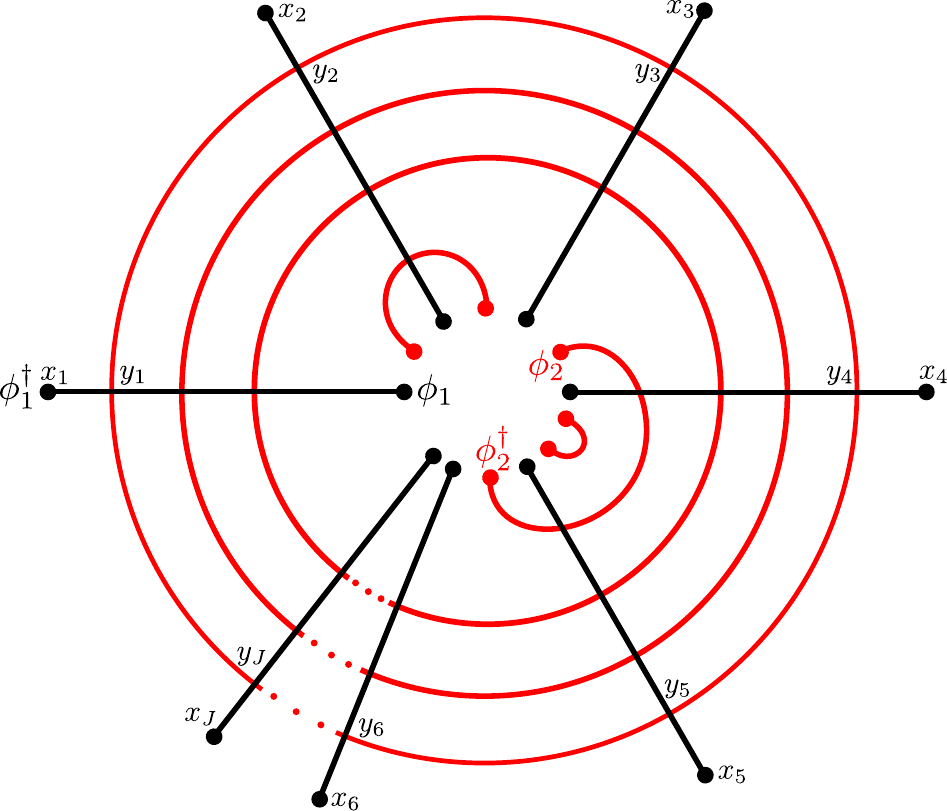}
	\caption{Feynman diagrams that contribute to the correlation function of an operator of the form $\tr\!(\d^m\phi_1^J(\phi_2\phi_2^\dagger)^n\dots)$ and $J$ $\phi_1^\dagger$ scalars are all of fishnet type -- made of the iterative wheel as shown on the diagram ~\cite{Nielsen:1970bc}. This structure can be resummed and leads to integrability~\cite{Zamolodchikov:1980mb}.}
	\label{fig:fishfig}
\end{figure}
One of the main features of the fishnet theory is the simple structure of the Feynman diagrams appearing in its perturbative expansion.
The Feynman diagrams which contribute to the correlation functions of the type \eq{wf}
are of the iterative fishnet type, after all $\phi_2$'s annihilate with $\phi_2^\dagger$, (see Fig.\ref{fig:fishfig}). The sum of these graphs can be represented, at least formally, as a simple geometric series $1/(1-\hat B)$. Here, the ``graph-building" operator, $\hat B$, is defined by its action on the wave function (\ref{wf}) that is a function of $J$ 4D variables~\cite{Gurdogan:2015csr}
\beq\la{gb}
\hat B\circ\varphi(\vec x_1,\dots,\vec x_J)\equiv \int\prod_{i=1}^Jd^4y_i\prod_{j=1}^J{\xi^2/\pi^2\over(\vec y_j-\vec y_{j+1})^2(\vec y_j-\vec x_j)^2}\,\varphi(\vec y_1,\dots,\vec y_J)\;.
\eeq
Acting with the operator $\hat B$ on the Feynman diagram in Fig.\ref{fig:fishfig} would create a new diagram with one extra wheel.     One can easily see that $\hat B$ is a self-adjoint operator with respect to the CFT norm (\ref{CFTnorm}). As exemplified in detail in \cite{Gromov:2018hut}, physical operators in the spectrum of the model correspond to wave functions that are stationary under the graph building operator
\beq\la{phys}
\hat B|\varphi_{phys}\rangle=|\varphi_{phys}\>\;.
\eeq

One can check explicitly that $\hat B$ commutes with all the conformal generators. This implies that the states which satisfy the physical condition (\ref{phys}) can be chosen to have a fixed conformal dimension and two spins, $\varphi_{\Delta,S_1,S_2}$. The physical spectrum of conformal dimensions $\Delta(S_1,S_2,\xi)$ is fixed from (\ref{phys}). For $J=2$ one finds two non-trivial states of the twist $t=2$ and $t=4$, \cite{Grabner:2017pgm}\footnote{For $J=2$ only the states with $S_2=0$ are possible.
	The twist is defined as $t=\Delta_0-S$, where $\Delta_0$ is the bare dimension. The two states correspond to $\tr\![\phi_1^2]$ and $\tr\![\phi_1\Box \phi_1]$ at weak coupling.}
\beq\la{disp}
\Delta_{t=2/4}=2+
\sqrt{(S_1+1)^2+1\mp 2\sqrt{(S_1+1)^2+4\xi^4}}\;.
\eeq

In addition to the states (\ref{phys}), there are also infinite towers of protected states. For example, at $J=2$ these are primary states with the twist bigger than four,  $\tr\!(\phi_1\Box^n\phi_1)$, $n>1$. The anomalous dimensions of all of these states vanish in the planar limit and they decouple from the interacting dilatation operator.\footnote{There are also {\it logarithmic} states, whose two-point function cannot be diagonalized~\cite{Gromov:2017cja}. Most likely all these states belong to the same Jordan cell of the dilatation operator as some of the protected states. We assume these states decouple from all non-trivial physical states, which follows from the condition \eq{phys}.} Hence, in this paper we will focus on the non-protected operators. In the next section we review the strong coupling dual of the physical states (\ref{phys}).

\subsection{The fishchain model}

In this section we describe the {\it fishchain} model, which was derived as a holographic dual of the planar fishnet theory in \cite{Gromov:2019aku} for the ${\mathfrak u}(1)$-sector. At strong coupling $\xi\to\infty$,
this model becomes an accurate description of the fishnet theory. 
The classical model is a chain of
$J$ classical point-like scalar particles that propagate on the lightcone $X_i^2=0$ of ${\mathbb R}^{1,5}$. Its dynamics is described by the 
discretised Nambu-Goto type action
\beq\la{Lup}
S=\xi\int\! L\, dt\ ,\qquad L=2 J\(
\prod_{i=1}^J\frac{\dot X_i^2}{-2X_i\cdot X_{i+1}}
\)^{\frac{1}{2J}}
\eeq
where $X_i^{M=-1,\dots,4}$ are flat embedding coordinates, \cite{Dirac:1936fq}.

Importantly, $\xi$ is the square root of the `t Hooft coupling in the fishnet model. Since it stands in front of the action, it also plays the role of $1/\hbar$ in the fishchain description. This ensures that the fishchain model is classical at strong coupling, as is expected from a holographic dual. 
This is analogous to the situation in ${\cal N}=4$ SYM theory. There, the worldsheet theory for a string moving in $AdS_5\times S^5$ becomes classical at strong coupling for those local single-trace operators whose dimensions scale as $\Delta\sim\sqrt\lambda$ at strong coupling.

The action (\ref{Lup}) is invariant under a global conformal symmetry. It acts by a simultaneous $SO(1,5)$ rotations at all $J$ coordinates $X_i$. The action is also invariant under two types of gauge symmetries. One is a time reparametrization, $t\to f(t)$, and the other is an independent rescaling at each site, $X_i(t)\to g_i(t)X_i(t)$. 

One way to fix the local rescaling gauge symmetry is to pick a slice of embedding space. Every slice gives a different conformally flat four-dimensional spacetime. For example, the {\it Poincar\'e} slice 
\beq\la{flat}
X_i^+=X_i^{-1}+X_i^0=1\ ,\qquad x_i^\mu=X_i^\mu
\eeq
takes us back to the flat space that was our starting point in \cite{Gromov:2019aku}. In this paper, we find it more useful to work in a covariant gauge. Namely, a gauge that commutes with all the $SO(1,5)$ global symmetries. One natural covariant gauge is $\dot X_i^2=m^2$, where $m^2$ is a constant. This is a suitable covariant gauge-fixing condition because $\dot X_i^2$ transforms non-trivially as $\dot X_i^2(t)\to g_i^2(t)\dot X_i^2(t)$.

To fix that gauge, we first follow \cite{Gromov:2019aku} and switch to the Polyakov form of the action  
\beq\la{actbetas}
L=\sum_i 
\[\frac{\alpha_i\dot X_i^2}{2}
+ \prod_k \(-\alpha_k X_k.X_{k+1}\)^{-\frac{1}{J}}+\eta_i X_i^2\]\;,
\eeq
where  $\alpha_i$'s are auxiliary fields. They transform covariantly under the two types of gauge symmetries. It is convenient to fix the gauge $\alpha_i=1$. The corresponding Virasoro type constraints are telling us that the energy density is zero along the chain of particles
\beq\la{Virasoro}
\dot X_k^2=2\prod_i \(-X_i. X_{i+1}\)^{-\frac{1}{J}}\equiv {\cal L}\ ,\qquad k=1,\dots,J\;.
\eeq

Because we had $J+1$ gauge symmetries and only $J$ gauge fixing conditions $\alpha_i=1$, we still remain with one gauge symmetry. It acts by a simultaneous time reparametrization and rescaling $t\to f(t),\;X_i\to X_i/\sqrt{f'},\;\eta_i\to \eta_i\,{f'^2}$. The covariant gauge divulged above corresponds to fixing $\cL=m^2$ to be a constant.\footnote{In \cite{Gromov:2019aku} we chose a different covariant gauge where $\sum_i\eta_i=J$.}

\begin{figure}
	\centering
	\includegraphics[scale=1.1]{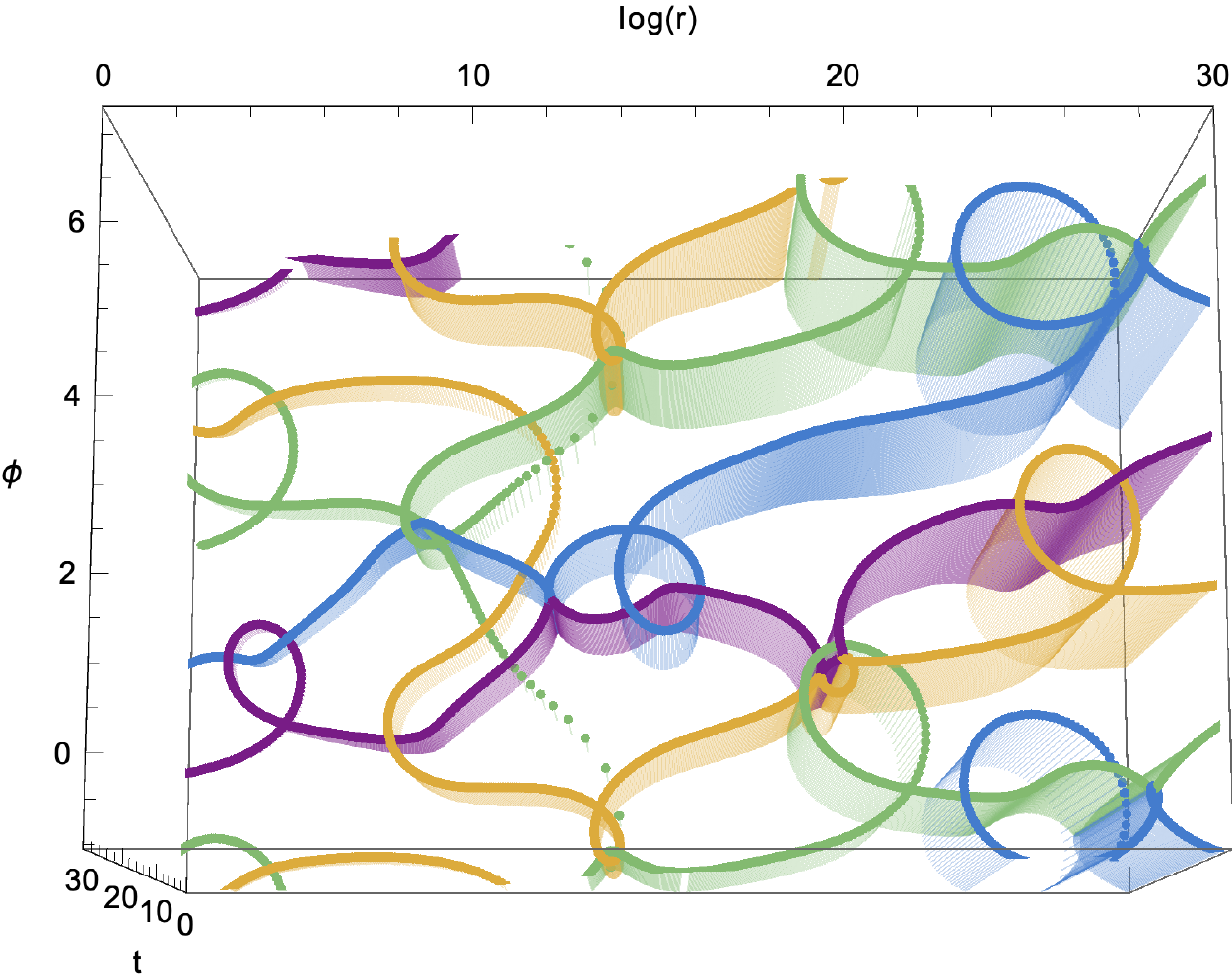}
	\caption{Dynamics of the fishchain consisting of $J=4$ particles (denoted by different colours). The initial conditions are restricted to a $2D$ plane after projection on the boundary. The plane is parametrized in the radial parameters by the angle $\phi$ and the radius $r$. The third axis corresponds to the time. The drift in $\phi$ is due to a non-zero total angular momentum $S$ and the exponential expansion in $r$ is due to $\Delta>0$.}
	\label{fig:particles}
\end{figure}
The equation of motion that follows from the corresponding gauge fixed action is
\beq\la{eomX}
\ddot X_i=2\eta_iX_i-{m^2\over2}\({X_{i+1}\over X_{i+1}.X_i}+{X_{i-1}\over X_i.X_{i-1}}\)\;.
\eeq
The dynamics encoded into these equations is quite non-trivial as one can see in Fig.~\ref{fig:particles}.
The term with $\eta_i$ is responsible for keeping the motion on the lightcone. 
A covariant form of this equation, where this longitudinal mode decouples, is
\beq\la{qcons}
\dot { q}_i=\frac{m^2}2 ({ j}_{i+1}-{ j}_i)\qquad\text{where}\qquad { j}_i^{MN}=2{X_{i-1}^{[M}X_i^{N]}\over X_{i-1}.X_i}
\eeq
and
\beq\la{qidef}
q_i^{M\,N}=\dot X_i^MX_i^N-\dot X_i^NX_i^M
\eeq
is the local charge density at the $i$'th site. The total $SO(1,5)$ symmetry charge is given by
\beq\la{totalcharge}
Q^{N\,M}=\xi \sum_i q_i^{N\,M}\equiv \xi {\cal Q}^{NM}\;.
\eeq
In particular for the highest weight solutions, i.e. for the classical solutions with the initial conditions chosen so that $Q^{NM}$ is block-diagonal, we have $Q^{-1,0}=i\Delta,\; Q^{12}=S_1$ and $Q^{34}=S_2$. The global charge $Q^{NM}$ is not the only conserved quantity in this theory. Due to the classical integrability of this model, which we describe in section \ref{sec6}, one can construct in total $\sim 4J$ non-trivial integrals of motion.

The starting point for the quantization, which we perform in the next section, is the Hamiltonian which is reminiscent of that of the conformal gauge fixed Polyakov action. It is obtained from \eq{actbetas} with $\alpha_i=1$ in the standard way  
\beq\la{Hclassical}
P_i^M=\dot X_i^M\qquad\Rightarrow\qquad
H=\sum_j\[{1\over2}P_i^2-\prod_k(-X_k.X_{k+1})^{-{1\over J}}\]\;.
\eeq
The list of the primary constraints is
\beq
X_i^2=0\ ,\qquad P_i^2=m^2\;\;,\;\;i=1,\dots,J\;.
\eeq
Since $H$ generates a gauge transformation, namely $t$-time translations, the physical solutions have zero energy $H=0$.

\section{Canonical quantization}\la{sec3}
In this section we describe in detail the procedure and the result of the canonical quantization of the fishchain model. 
Then in section \ref{sec4} we give some detailed examples of our construction and compare the results of the quantization of our model with  known CFT predictions.

\subsection{Warm-up example: point-like particle on a sphere}
As a warm-up exercise, we begin from a simple model with  well-known properties -- a point-like particle on a sphere. The process of canonical quantization of this model is very similar to that of the fishchain, and is very instructive for the  next section. 

\subsubsection{Classical theory and constraints}

We describe a free particle on the unit sphere as a system with a constraint:
\beq
L=\frac{m\dot{ x}^2}{2}+\lambda( x^2-R^2)\ ,
\eeq
where $ x$ is a $d$ dimensional vector and $\lambda$ is a scalar Lagrange multiplier enforcing the $ x^2=R^2$ constraint. The naive Hamiltonian of the system is 
\beq\la{Hnaive}
H=\frac{ p^2}{2m}-\lambda( x^2-R^2)\;.
\eeq
Following the standard procedure we have to identify all the further constraints which could arise from the primary constraint $\varphi_1=x^2-R^2$ and modify the initial Hamiltonian, so that on the constraints shell it is equivalent to \eq{Hnaive}. We have one secondary constraint
\beq\la{phi2}
\dot\varphi_1=\{\varphi_1,H\}_{PB}=\frac{2}{m}( x. p)\equiv \frac{2}{m}\varphi_2\;,
\eeq
where the Poisson bracket is defined in the standard way, as in a non-constrained system
\beq
\{f,g\}_{PB}=\sum_{i=1}^d\(\frac{\d f}{\d x_i}\frac{\d g}{\d p_i}-\frac{\d g}{\d x_i}\frac{\d f}{\d p_i}\)\;.
\eeq
Next, we should try to modify the Hamiltonian to prevent the appearance of the further constraints. A nice way to do this is to introduce the angular momentum associated with the $O(d)$ symmetry 
\beq
q^{mn}=x^mp^n-p^mx^n
\eeq
and notice that ${\rm tr}\;q^2=-\sum\limits_{n,m}(q^{nm})^2=-2 x^2p^2+2 (x.p)^2\simeq -2 R^2 p^2$, so that we can replace the naive Hamiltonian by
\beq\la{Hqq}
\tilde H=-\frac{{\rm tr}\;q^2}{4R^2m}\;.
\eeq
The modified Hamiltonian $\tilde H$ coincides with the initial one on the constraint shell.
Since $q^{mn}$ generates rotations, their Poisson bracket with any scalar combination is zero, which includes $\varphi_1$ and $\varphi_2$. Thus, there are no further constraints generated by the Poisson bracket with the Hamiltonian. At the same time, the Poisson bracket between $\varphi_1$ and $\varphi_2$ is non-zero, even on-shell
\beq
\{\varphi_1,\varphi_2\}=2x^2\simeq 2 R^2\;.
\eeq
This means that $\varphi_1$ and $\varphi_2$ are the {\it second class} constraints. The existence of the second class constraints implies that we cannot quantize the model by simply replacing the Poisson bracket with the commutator. Obviously, we would run into a contradiction by imposing both $\hat \varphi_1$ and $\hat \varphi_2$ to vanish on the physical states. The standard way to proceed is to replace the Poisson bracket with the Dirac bracket
\beq\la{DB0}
\{f,g\}_{DB}=\{f,g\}_{PB}+\sum_{a,b}M_{ab}\{f,\varphi_a\}_{PB}\{\varphi_b,g\}_{PB}
\eeq
where $M_{ab}$ is an anti-symmetric matrix, fixed by the condition that $\{f,\varphi_a\}_{DB}=0$ for any $f(x,p)$. In our case we find
\beq\la{DB}
\{f,g\}_{DB}=\{f,g\}_{PB}-\frac{\{f,\varphi_2\}_{PB}\{\varphi_1,g\}_{PB}-\{f,\varphi_1\}_{PB}\{\varphi_2,g\}_{PB}}{2R^2}\;.
\eeq
The quantization is then obtained by replacing the observables by operators and the Dirac bracket by the commutators.

\subsubsection{Quantization}

To build the quantum algebra of observables, the first step is to represent $x_i$ and $p_i$ as operators acting on a Hilbert space.
This, however, is not always a straightforward task. The problem is that the Dirac bracket between $x_i$ and $p_i$ is modified in comparison to the Poisson bracket due to the presence of the second class constraints
\beqa\la{xpdb}
\{x_i,x_j\}_{DB}=0\ ,\quad
\{p_i,p_j\}_{DB}=\frac{p_i x_j-p_j x_i}{R^2}\ ,\quad
\{x_i,p_j\}_{DB}=\delta_{ij}-\frac{x_i x_j}{R^2}\;.
\eeqa
Under the canonical quantizaiton we replace $\{\cdot,\cdot\}_{DB}\to \frac{1}{i\hbar}[\cdot,\cdot]$ to get
\beqa\la{xp1}
&&\frac{1}{i\hbar}[\hat x_i,\hat x_j]=0\ ,\\
\la{xp2}&&\frac{1}{i\hbar}[\hat p_i,\hat p_j]={\hat p_i \hat x_j-\hat p_j\hat x_i\over R^2}\ ,\\
\la{xp3}&&\frac{1}{i\hbar}[\hat x_i,\hat p_j]=\delta_{ij}-{\hat x_i\hat x_j\over R^2}\;.
\eeqa
It is clear that the standard representation where $\hat p_i\to -i\hbar \d_{x_i}$ is not compatible with \eq{xp3}. The good news, however, is that $\hat x_i$'s do commute with each other and we can diagonalize them simultaneously. This allows us to introduce the standard coordinate representation for the Hilbert space in terms of the functions of $d$ variables $\Psi(x_1,\dots,x_d)$. Now, to find the representation for $\hat p_i$, we introduce the notation $\hat k_i\equiv -i\hbar \d_{x_i}$. At this point there is no clear relation between $\hat p_i$ and $\hat k_i$.
However, we can always define operators $\hat o_i$ such that 
\beq
\hat p_i=\hat k_i-\frac{(\hat k.\hat x)\hat x_i}{R^2}+\hat o_i\;.
\eeq
By plugging this definition into \eq{xp3}, and using that $\frac{1}{i\hbar}[\hat x_i,\hat k_j]=\delta_{ij}$
we see that $[\hat o_i,\hat x_j]=0$ meaning that the operators $\hat o_i$ are functions of $\hat x_i$.
The only function of $\hat x_i$, which has the right properties under the $O(d)$ symmetry is $\hat x_i$ itself (assuming that the constraint $\hat x^2=R^2$ is already imposed). So we conclude
\beq\la{eqp}
\hat p_j=\hat k_j-\frac{(\hat k.\hat x)-i\hbar c}{R^2}\hat x_j\;,
\eeq
where $c$ is some unknown constant. We see that the term with $(\hat k.\hat x)$ is responsible for rescaling $\hat x$ back to the sphere and the constant $c$ represents an ordering ambiguity between $\hat k$ and $\hat x$. It is left to impose \eq{xp2}, but one can check that it is automatically satisfied.

\subsubsection{Hilbert space}
We have to examine the constraints and describe the physical sub-space of the whole Hilbert space.
First, we note that the constraint $\hat \varphi_1=\hat x^2-R^2$ can be consistently imposed. For this we find that
\beq\la{comphi1}
\frac{1}{i\hbar}[\hat \varphi_1,\hat x_i]=0\ ,\qquad
\frac{1}{i\hbar}[\hat \varphi_1,\hat p_i]=-\frac{\hat \varphi_1\hat x_i}{R^2}\;.
\eeq
This means that we can restrict ourselves to the sub-space ${\cal H}$, which consists of the functions $\Psi(\vec x)$ such that
\beq\la{subs}
\hat \varphi_1|\Psi\>=0\quad\Leftrightarrow\quad\Psi(x)=R\,\delta( x^2-R^2)\times\psi(\vec x)\;.
\eeq
One can then say that the non-trivial part of the  wave-function $\psi$ only depends on the coordinates on the sphere. Acting with any combination of $\hat x_i$ and $\hat p_i$ will not lead outside this subspace due to \eq{comphi1}. This is opposite to the action of $\hat k_i$, which does lead outside ${\cal H}$. The commutation relations \eq{comphi1} imply that we can consistently restrict all the observables to the space of functions on the sphere. 

The constraint $\hat \varphi_2$ can be explicitly written as
\beq\la{phi2q}
\hat \varphi_2=\hat { p}.\hat { x}=
\frac{i\hbar c \hat x^2}{R^2}+(\hat k.\hat x)\(1-\frac{\hat x^2}{R^2}\)\approx i\hbar c
\eeq
where the $\approx$ sign means equality on the functions that satisfy the $\hat\varphi_1$ constraint. Hence, to enforce this constraint it is sufficient to fix $c=0$. Note, however, that one could take another ordering $\hat x.\hat p\approx i\hbar(c+d-1)$, which would imply $c=1-d$, or any other ordering. In either case the constraint $\hat \varphi_2$ is just a constant on ${\cal H}$ and the exact value of $c$ would correspond to a particular choice of the ordering.
In what follows we will not set $c$ to any particular value.
We will see that it disappears from all relevant physical quantities.

\paragraph{Scalar product.}
Note that the subspace ${\cal H}$ consists of, strictly speaking, non-normalizable wave-functions, due to the $\delta$-function prefactor in \eq{subs}. At the same time,
the regularised scalar product is naturally defined as
\beq
\langle \psi_1 |\psi_2\rangle\equiv \int d^d x\,
R\,\delta(\vec x^2-R^2)\,\bar \psi_1(\vec x)\psi_2(\vec x)\;.
\eeq

\subsubsection{Spectrum and stationary wave-functions}
Like in the classical case, the quantum Hamiltonian has the following two equivalent on-shell expressions
\beq\la{ppC2}
\hat H=\frac{\hat p^2}{2m}\approx -\frac{{\rm tr}\,\hat q^2}{4R^2 m}-\hbar^2\frac{c(c+d-1)}{2R^2 m}
\eeq
which is now corrected at the quantum level by the last term, which vanishes exactly when either $\hat x.\hat p$ or $\hat p.\hat x$ is set to zero. We  use $\approx$ for the expressions valid on the subspace ${\cal H}$ i.e. on the functions satisfying \eq{subs}. 

Even though the commutator of $\hat x_i$ and $\hat p_i$ is modified, the commutator of the components of the charge density $\hat q^{nm}$ satisfy the standard ${\frak{so}}(d)$ algebra
\beq
\frac{1}{i\hbar}[\hat q^{ab},\hat q^{ce}]=\hat q^{ac}\delta^{be}-
\hat q^{bc}\delta^{ae}-
\hat q^{ae}\delta^{bc}+
\hat q^{be}\delta^{ac}\;.
\eeq
In the language of the ${\frak{so}}(d)$ algebra the Hamiltonian is proportional to the quadratic Casimir operator. This means that its eigenvalues are fixed by   representation theory. In particular, for representations given by  symmetric traceless tensors with $S$ indexes the eigenfunctions of the Hamiltonian should be
\beq
C_2\equiv -\frac{1}{2\hbar^2}{\rm tr}\,\hat q^2 =  S(S+d-2)\;.
\eeq
The corresponding wave functions are the traceless symmetric tensors which can be conveniently parametrised by a light-like (complex) vector $\vec n$ as
\beq
\psi_{S,\vec n}(\vec x)=(\vec x.\vec n)^S\;.
\eeq
Note that there are no other irreducible tensors one can build out of one unit vector $\vec x/R$.
Thus we conclude the spectrum is
\beq
E_S=\hbar^2\frac{ S(S+d-2)-c(c+d-1)}{2R^2m}\;.
\eeq

\subsection{Quantization of the fishchain model}\la{sec:fcq}
After recalling the general procedure of the canonical quantization of the system with constraints we can proceed to the quantization of the fishchain model, which carries a number of similarities with the example above.

In this section we use the notation $D$ for the dimension of the embedding space and $d\equiv D-2$ is the dimension on the boundary. Even though we are mostly interested in $D=6$ and $d=4$, we try to keep $D$ arbitrary wherever  possible.
\subsubsection{Algebra of constraints}
As in the previous section we have to list all primary constraints, then generate secondary constraints with a possibility of modifying the Hamiltonian by adding any linear combinations of the constraints. After that we have to identify second class constraints (whose Poisson bracket is non-zero on-shell) and introduce the Dirac bracket, which can then be quantized.

Firstly, we have the Hamiltonian constraint (\ref{Hclassical}) that we repeat here for convenience 
\beq
\varphi_H=H=\sum_i{P_i^2\over2}-J\prod_k(-X_k.X_{k+1})^{-{1\over J}}
\;.
\eeq
Secondly, we have $2J$ more 
primary constraints defined by
\beq
\varphi_{i,X}=X_i^2\ ,\qquad \varphi_{i,P}=\frac{1}{2}(P_i^2-m^2)\;.
\eeq
Next, we have to work out the secondary constraints, by computing the Poisson brackets of the primary ones with the modified Hamiltonian 
\beq
H'=H-\sum_i (u_{i,X}\varphi_{i,X}+u_{i,P}\varphi_{i,P})\;.
\eeq
that is equivalent to $H$ on the $\varphi_{i,X}$ and $\varphi_{i,P}$ constraint shell. Here, the $u$'s are arbitrary functions of $X$ and $P$. The Poisson bracket is defined in the standard way $\{X_i^N,P_j^M\}_{PB}=\delta_{ij}\eta^{NM}$.
For $\varphi_{i,X}$ we have 
\beq\la{HphiX}
\{\varphi_{i,X},H'\}_{PB}=2(1-u_{i,P}) X_i.P_i\;.
\eeq
Unless $u_{i,P}=1$, it leads to a new constraint, analogously to the \eq{phi2} from the previous section
\beq
\varphi_{i}\equiv X_i.P_i\;.
\eeq
Together with $\varphi_{i,X}$ and $\varphi_{i,P}$ we have $3J$ constraints, which form the following algebra
\beq\la{phii}
\{\varphi_{j,X},\varphi_{j}\}_{PB}=2\varphi_{j,X}\ ,\quad
\{\varphi_{j},\varphi_{j,P}\}_{PB}=2\varphi_{j,P}+m^2\ ,\quad
\{\varphi_{j,P},\varphi_{j,X}\}_{PB}=2\varphi_{j}\;.
\eeq
Then, we should also check that $\{\varphi_{i,P},H'\}_{PB}$
does not generate any further constraints. For that we should allow the secondary constraint $\varphi_{i}$ \eq{phii} to be included in the possible modification of the Hamiltonian 
\beq
H''=H-\sum_i (u_{i,X}\varphi_{i,X}+u_{i,P}\varphi_{i,P}+u_{i}\varphi_{i})\;.
\eeq
Note that with this further modification we still have $\{\varphi_{i,X},H''\}\simeq 0$, where the $\simeq$ sign means equality on functions that satisfy the constraint. Next, from $\dot \phi_{i,P}$ we have 
\beqa
\{\varphi_{i,P},H''\}_{PB}
&\simeq& 
\[2u_{i}-\frac{P_i.X_{i+1}}{X_i.X_{i+1}}+\frac{P_i.X_{i-1}}{X_i.X_{i-1}} \]\frac{m^2}{2}\;.
\eeqa
To prevent further constraints we can fix the parameter
\beq\la{ui}
u_{i}=\frac{P_i.X_{i+1}}{2X_i.X_{i+1}}+\frac{P_i.X_{i-1}}{2X_i.X_{i-1}}\;.
\eeq
This also rules out the option of setting $u_{i,P}$ to one in (\ref{HphiX}) because then we will keep generating more and more constraints.

Finally, we have to make sure that $\dot\varphi_{j}=\{\varphi_{j},H''\}$ stays on the constraint shell.
It is easy to see that this fixes $u_{i,P}=0$, due to  
\beq
\{\varphi_{j},H''\}_{PB}\simeq -m^2u_{j,P}\;.
\eeq
After that there are no further constraints and the Hamiltonian takes the form
 
\beq
H''=\sum_j{P_i^2\over2}-J\prod_k(-X_k.X_{k+1})^{-{1\over J}}
-\sum_i (u_{i,X}\varphi_{i,X}+u_{i}\varphi_{i})\;,
\eeq
where $u_i$ are fixed by \eq{ui} and $u_{i,X}$ remains arbitrary.
This clear difference between $\varphi_{i,X}$ and the other constraints is due to the fact that it is a {\it first class} constraint, whereas $\varphi_{i,P}$ and $\varphi_i$ are {\it second class} constraints. This can also be seen from their Poisson bracket \eq{phii}, which is non-zero on-shell. The existence of the second class constraint implies that we will have to introduce the Dirac bracket before proceeding to quantization, just like in the previous section. We will proceed to this step in the next section. Here we simplify the Hamiltonian.

Note that we have the freedom of adding to the Hamiltonian terms that are quadratic in constraints. This is a freedom in the canonical quantisation procedure that does not affect the result. Indeed, such a modification would not change any of the Poisson bracket relations on-shell, but may allow us to simplify the result. 
In the search for a simpler Hamiltonian we can look for  inspiration from the previous section, where we got a nice expression in terms of the square of the global charge \eq{Hqq}. This suggests using the charge density \eq{qidef}
 
\beq\la{cd}
q_i^{M\,N}=P_i^MX_i^N-P_i^NX_i^M
\eeq
as a building block. By observing that
\beq\la{qq}
(q_i^2)^{NM}=
-m^2X_i^N X_i^M-2\varphi_{i,P}X_i^N X_i^M+
\varphi_i(P_i^NX_i^M+X_i^N  P_i^M)-\varphi_{i,X}P_i^N P_i^M\;.
\eeq
we expand $\tr\prod_i q_i^2$ in powers of the constraints. We get
\beqa\la{trqq}
\nn\tr\( \prod_i \frac{q_i^2}{2}\)-1&=&\(\frac{m^2}{2}\)^J\prod_i\(-X_i.X_{i+1}\)-1\\&+&
\frac{2}{m^2}\sum_i(\varphi_{i,P}+v_i \varphi_{i,X})
-\frac{2}{m^2}\sum_i\varphi_i u_i+{\cal O}(\varphi^2)
\eeqa
where $u_i$ is exactly like the one we found before in 
\eq{ui} and the precise form of $v_i$ is not so important for what follows.
Next, by using the exact identity
\beq
\frac{m^2}{2}=\prod_k(-X_k.X_{k+1})^{-{1\over J}}-\frac{1}{J}\(\sum_i\varphi_{i,P}-\varphi_H\)
\eeq
multiplying both parts by
$\prod_k(-X_k.X_{k+1})^{{1\over J}}$
and raising both parts to the power $J$ 
we obtain, up to terms quadratic in the constraints
\beq
\(\frac{m^2}{2}\)^J\prod_k(-X_k.X_{k+1})-1=-\frac{2}{m^2}\(\sum_i\varphi_{i,P}-\varphi_H\)+{\cal O}(\varphi^2)
\eeq
which together with \eq{trqq}
implies that 
\beq\la{fcH}
H''=\frac{m^2}{2}H_q-\sum_i u_{i,X} \varphi_{i,X}+{\cal O}(\varphi^2)\ ,\qquad
H_q\equiv \tr\( \prod_i \frac{q_i^2}{2}\)-1\;.
\eeq

Thus we have shown that up to an irrelevant constant multiplier 
$H''$ and $H_q$ are equivalent.
One can also move the $\sum_i u_{i,X} \varphi_{i,X}$ terms into the definition of $H_q$, however, this will not make any difference as this is the first class constraint, which Poisson-commutes with all other constraints. 
We will be using $H_q$ in what follows.

Let us note that by using $H_q$ it becomes obvious that there are no further constraints generated from $\{\varphi,H_q\}_{PB}$. Indeed, the Poisson bracket of $q_i$ generates the rotation at the $i$-th cite of the chain. Since all our constraints are scalar combinations of $X_i^N$ and $P_i^M$ they must commute with $H_q$ by construction. 

\subsubsection{Dirac bracket}
In the previous section we derived the system of constraints which contains $2J$ second class constraints $\varphi_{i,P}$ and $\varphi_{i}$, which do not Poisson-commute with each other on-shell. To fix this problem one should introduce the Dirac bracket, as we already demonstrated above. The modification is very similar to what we have done for the case of a single particle on a sphere \eq{DB} and is given by
\beq\la{DB2}
\{f,g\}_{DB}=\{f,g\}_{PB}-\sum_i\frac{\{f,\varphi_{i,P}\}_{PB}\{\varphi_i,g\}_{PB}-\{f,\varphi_i\}_{PB}\{\varphi_{i,P},g\}_{PB}}{m^2}\;,
\eeq
which is designed so that
\beq
\{\varphi_{i,P},g\}_{DB}=0\ ,\qquad
\{\varphi_{i},g\}_{DB}=0
\eeq
for any function $g$ on the phase space.
The main difference with \eq{DB} is that momentum and space coordinates get interchanged. As a consequence the Dirac bracket
between the $X$'s and $P$'s is the same as \eq{xp1}-\eq{xp3}
with $P$ playing the role of $x$ and $X$ playing the role of $p$:
\beqa\la{xp_db}
\{X_i^M,P_j^N\}_{DB}&=&\delta_{ij}\[\eta^{MN}-\frac{1}{m^2}P_i^MP_i^N\]\;,\\
\{P_i^M,P_j^N\}_{DB}&=&0\;,\\
\{X_i^M,X_j^N\}_{DB}&=&\delta_{ij}\frac{P_i^MX_i^N-P_i^NX_i^M}{m^2}\;.
\eeqa
We can now proceed to the canonical quantization by promoting the Dirac bracket into the commutator.

\subsubsection{Quantization}
Following the standard procedure we now replace the Dirac bracket with a commutator, where the role of $\hbar$ is played by $\frac{1}{\xi}$ as we discussed in the introduction
\beqa
\nn[\hat X_k^M,\hat P_j^N]&=&\frac{i\delta_{kj}}{\xi}\[\eta^{MN}-\frac{1}{m^2}\hat P_k^M\hat P_k^N\]\;,\\
\la{comXP0}{}[\hat P_k^M,\hat P_j^N]&=&0\;,\\
{}[\hat X_k^M,\hat X_j^N]&=&\frac{i\delta_{kj}}{\xi}\frac{\hat \nn P_k^{M}\hat X_k^{N}-\hat P_k^{N}\hat X_k^{M}}{m^2}\;.
\eeqa
Now we have to build a representation of the operators $\hat X_i^M$ and $\hat P_i^M$. In general, building the representation of the algebra defined by \eq{comXP0} is a non-trivial problem.
However, in our case we notice that the algebra is exactly the same as in the case of the particle on a sphere, with $\hat P_i$ playing the role of $\hat x$ and $\hat X_i$ playing the role of $\hat p$. Following the procedure developed for the point-like particle we have to introduce a new set of operators $\hat Y_i$, which commute with $\hat P_i$ in the standard way
\beqa\la{comXP}
[\hat Y_k^M,\hat P_j^N]=\frac{i}{\xi}\delta_{kj}\eta^{MN}\ ,\qquad[\hat Y_i^M,\hat Y_j^N]=0
\eeqa
and then the operators $\hat X_i$ are given by
\beq\la{XtoY}
\hat X_i^M=-\frac{i c_i}{m^2 \xi }\hat P_i^M+\hat Y_i^M-\frac{1}{m^2}\hat P_i.\hat Y_i\; \hat P_i^M\;.
\eeq
One can check that all the commutation relations \eq{comXP0} are automatically satisfied. Thus we can represent the Hilbert space as a space of functions $\Psi(Y^N_1,Y^N_2,\dots,Y^N_J)$ on which $\hat Y_i^N$ acts as a multiplication by $Y_i^N$
and $\hat P_i^M=-\frac{i}{\xi}\eta^{MK}\d_{Y_{i}^K}$ in the usual way, and $\hat X_i$ is a non-trivial operator acting according to \eq{XtoY}.

\paragraph{Quantum constraints.} 
To proceed, we have to write the quantum version of the constraints. The constraints $\varphi_{j,P}$ and $\varphi_j$ are analogous to those for the case of the sphere and can be treated similarly. First, we impose the $\varphi_{j,P}$ constraint on the Hilbert space
\beq\la{P2constraint}
(\hat P_j^2-m^2)\Psi=0\;.
\eeq   
Notice that on the constraint \eq{P2constraint}
one can replace $\hat Y_j$ by $\hat Y_j+\alpha_j \hat P_j$,
with any constant $\alpha_j$ without changing $\hat X_j$ in the relation \eq{XtoY}. It indicates that $\Psi$,  as a function of $\hat Y_i$, contains non-physical information and can be further reduced. In other words, the constraint (\ref{P2constraint}) fixes the dependence of the wave function under translations of $\hat Y$ in the direction of $\hat P$ and is related to the bulk equation of motion. In addition, the radial direction $R_i^2=-Y_i^2$ is not physical and should decouple from the wave function. 
This will be demonstrated below  in a more formal way.

Now let us process the constraint $\varphi_j$. Using $\eq{XtoY}$ on the subspace (\ref{P2constraint}) we have 
\beq
\hat X_j.\hat P_j=\frac{c_j-D}{i\xi}+\(\hat P_j.\hat Y_j-\frac{c_j}{i\xi}\)\(1-\frac{\hat P_j^2}{m^2}\)\approx \frac{c_j-D}{i\xi}
\eeq
and similarly $\hat P_j.\hat X_j\approx \frac{c_j-1}{i\xi}$. We see that in order to impose the $\varphi_j$ constraint we have to
choose a particular value for $c_j$ in (\ref{XtoY}), which depends on the ordering.
This is in complete analogy with \eq{phi2q} for a particle on a sphere. We can reverse the logic and take $c_j$ arbitrary,
and interpret it as a quantum correction to the classical $\varphi_j=0$ condition $\hat\varphi_j\approx \frac{c-D}{i\xi}={\cal O}(\hbar)$.
Following this logic we should also allow for a quantum correction to the $\varphi_{i,X}$ constraint.
To investigate this possibility
let us rewrite $\hat X_j^2$ in terms of $\hat P_j$ and $\hat Y_j$
in the way similar to \eq{ppC2}
\beq
\hat X_j^2\approx
\frac{\hat {\bf C}_{2,j}-(c_j-D)(c_j-1)}{m^2\xi^2}
\eeq
where $\hat {\bf C}_{2,j}$ is the quadratic Casimir operator of the conformal group on a single site
\beq\la{Cinq}
\hat {\bf C}_{2,j} \equiv \frac{\xi^2}{2}
\hat q^{MN}_j\hat q^{M'N'}_j\eta_{NN'}\eta_{MM'}=-\frac{\xi^2}{2}{\rm tr}\;\hat q^2_j\ ,
\eeq
where we use that the local charge density can be defined equivalently with $\hat X_i$ or with $\hat Y_i$, as the extra terms in \eq{XtoY} cancel and the result is the same 
\beq\la{qXP}
\hat q^{NM}_j\equiv \hat X_j^N \hat P_j^M-\hat X_j^M \hat P_j^N=\hat Y_j^N \hat P_j^M-\hat Y_j^M\hat P_j^N\;.
\eeq
The local operator \eq{Cinq} commutes with all constraints and with the Hamiltonian, and thus should have a physical
interpretation. Its eigenvalue  is a kinematic number that is dictated by the symmetry. At this point that number cannot be fixed and reflects the ambiguity in the quantization procedure.
However, it has a clear physical meaning. The particles in the fishchain model are essentially the $\phi_1$ scalars in the original CFT. These scalar fields are protected operators and carry dimension $\Delta_i=1$. From that it is natural to assign the dimension $\Delta_i=1$ to each of the particles constituting the fishchain. Note that classically there would be no difference for any value $\Delta_i\sim 1$ as the total dimension scales classically as $\xi\to\infty$ and the charges of the individual particles are negligible, which is why this information is missing in the classical limit.
Since the eigenvalue of the quadratic Casimir $\hat {\bf C}_2$ is given by  $\Delta_i(\Delta_i-D+2)+S(S+D-4)$ from the above argument we should set $\Delta_i=1$ and $S=0$  to obtain
\beq\la{quadratic}
{\bf C}_{2,j}=-3\frac{d^2}{16}\;.
\eeq
Below we  verify this natural prescription on a number of examples and also present the general proof that this prescription leads to the correct duality with the CFT side. 
Furthermore, this relation is consistent with the quantum integrability of the fishchain model as is discussed in section \ref{sec6}. 

\begin{figure}
	\centering
	\def\svgwidth{10cm}
	
	\ifpdf
	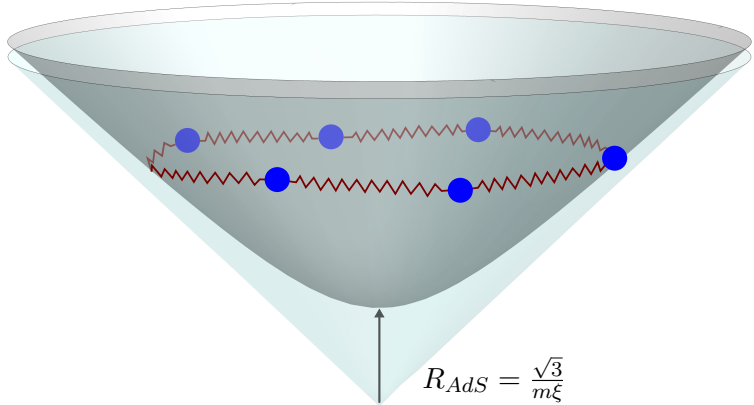
	\else
	\com{PDF-picture replacement}
	\fi

	\caption{Due to the quantum corrections we find an emergent non-zero AdS radius.}
	\label{fig:cones}
\end{figure}

Returning to the discussion of the $\varphi_{j,X}$ constraint, we notice that by setting $c_j$ to $1$ or to $D$, which would ensure $\hat\varphi_j=0$ for one of the most natural orderings, we get simply
\beq\la{XX}
\hat X_j^2\approx
-3\frac{d^2}{16m^2\xi^2}\;.
\eeq
This relation can be interpreted as an emergence of the non-zero quantum AdS radius due to  quantum corrections $R=\sqrt 3\frac{d}{4m\xi}$. In the classical limit $\xi\to\infty$ and $R\to 0$, which explains why the classical model lives on the lightcone $X^2_i=0$ (see Fig.\ref{fig:cones}). Below we discuss this crucial fact in more detail.

\paragraph{Quantum Hamiltonian.}
Finally, we have to examine the possibility of the appearance of the quantum corrections in the Hamiltonian $H_q$ itself.
Classically $H_q$ is given by \eq{fcH} which is built out of $(q_j^2)^{NM}$. These are $D\times D$ matrices with the following properties:
they are symmetric matrices (as a square of antisymmetric) and traceless,
i.e. they are irreducible symmetric tensors. However, at the quantum level these properties are lost. First, due to non-commutativity between the matrix elements of $\hat q_j^{NM}$
we have a corrected relation 
\beq
(\hat q_j^2)^{NM}-(\hat q_j^2)^{MN}=\frac{i\;d}{\xi}\hat q^{NM}_j\;.
\eeq
Furthermore, the trace is related to the quadratic Casimir \eq{quadratic}, which is non-zero
\beq\la{cas}
{\rm tr }\;\hat q_j^2=-\frac{2}{\xi^2}\hat{\bf C}_{2,j}\simeq
\frac{3d^2}{8\xi^2}\;.
\eeq
To recover the irreducibility we define the ``normal ordered" $\hat q^2_j$ as  
\beq\la{noq}
{:\hat q^2_j:}^{NM}\;\equiv\;
(\hat q^2_j)^{NM}-\frac{i\;d}{2\xi}\hat q_j^{NM}-\frac{3d^2}{8D\xi^2}\eta^{NM}
\;,
\eeq
which is again traceless and a symmetric tensor.
We use this natural prescription, which removes many possible ambiguities, in order to define the quantum Hamiltonian as
\beq\la{qH}
\hat H_q = 
\tr\( \prod_j \frac{:\hat q_j^2:}{2}\)-1\;.
\eeq

We will show in section~\ref{sec6} that this normal ordering procedure ensures  integrability at the quantum level.

\paragraph{Simplifying constraints.}
In order to describe in detail the physical Hilbert space we are going to give a clear interpretation of the constraints.
We understood that at each site of the chain we have to impose two conditions
\eq{P2constraint} and \eq{quadratic}. Namely
\beq
-\frac{1}{\xi^2}\Box_j\,\Psi=m^2 \Psi\ ,\qquad
\hat {\bf C}_{j,2}|\Psi\> = 
-3\frac{d^2}{16}|\Psi\>
\eeq
where $\Box_j$ is the D-dimensional d'Alembertian in $Y_{j}^M$ and $\hat {\bf C}_{j,2}$ is the quadratic Casimir operator built out of the $SO(1,D-1)$ charge densities via \eq{Cinq}.

We are going to show that the auxiliary coordinates $Y_j^M$ contain one extra unphysical direction, just like for the particle on a sphere. This  is the direction along the AdS radius $R_j^2\equiv - Y_j^2$. 
Using the identities
\beq\la{FR}
\hat Y_j^2\hat P_j^2=(\hat Y_j.\hat P_j)^2-i\frac{d}{\xi}\hat Y_j.\hat P_j
+\frac{\hat {\bf C}_{j,2}}{\xi^2}\ ,\qquad \hat Y_j.\hat P_j=-\frac{i}{\xi }R_j\d_{R_j}
\eeq
we obtain
\beq\la{dR}
\frac{1}{\xi^2}\Box_j\Psi=
\frac{1}{\xi^2R_j^2}\((R_j\d_{R_j})^2+{d}R_j\d_{R_j}-{\bf C}_{j,2}\)\Psi
=-m^2 \Psi\;.
\eeq
We see that the dependence of the wave function on $R_j$
is fixed by the equation \eq{dR}. By introducing the function $F(R_j)$, which solves the equation \eq{dR} the dependence on all $R_j$'s can be factored out
\beq\la{toAdS}
\Psi(Y_1,\dots,Y_J)=\psi(Z_1,\dots,Z_J)\times \prod_j F(R_j)\;,
\eeq
where we have introduced coordinates on a unit $AdS_{D-1}$ space, $Z_j^2=-1$ such that $Y_j^M=R_j\, Z_j^M$. Let us emphasize that the factor $F(R)$ is totally determined by the equation \eq{FR} (up to suitable boundary conditions), and does not contain any dynamical information. It is analogous to the
$\delta(x^2-R^2)$ factor for the case of the sphere.\footnote{The reason it is not $\delta$-function in our case is due to the change of roles between space and momentum. In fact the Fourier transform of the $\delta(r^2-R^2)$ factor is very similar to the $F(R)$ type of function.} The solutions to \eq{FR} are given by the Bessel function $F(R)=R^{\frac{2-D}{2}}J_{\frac{D-2}{4}}(m\xi R)$. Discarding this trivial factor we remain with a wave function on unit-radius $AdS$ space that is subject to only one constraint for each site
\beq\la{constr}
\hat {\bf C}_{i,2}|\psi\>= {\bf C}_2\,|\psi\>\;,
\eeq
where we use that the local charge $\hat q_i$ written in the coordinates $R_j,Z_j^N$ contains no explicit dependence on $R_j$, so the equation \eq{constr} is well-defined 
\beq\la{qZ}
\hat q^{MN}_j=-{i\over\xi}\(Y_j^N{\d\over\d Y_j^M}-Y_j^M{\d\over\d Y_j^N}\)=-{i\over\xi}\(Z_j^N{\d\over\d Z_j^M}-Z_j^M{\d\over\d Z_j^N}\)\;.
\eeq
The equation \eq{qZ} implies that both $\hat {\bf C}_{j,2}$ and $\hat H_q$ commute with $\hat R^2$ and take the same form (\ref{cas}), (\ref{qH}) when acting on $\psi$. 
In particular, $\hat {\bf C}_{i,2}$ is nothing but the covariant Laplacian on AdS and the remaining constraint \eq{constr} is simply the Klein-Fock-Gordon equation for a scalar field of mass $m^2\,R_{AdS}^2=-3d^2/16$, dual to the dimension-one scalar $\phi_1$ of the fishnet model.

To summarise the main results of this section: the wave function of the whole chain lives in the tensor product of a free massive scalar on AdS parametrized by $Z_i^2=-1$. The interaction between the particles comes from the one remaining Hamiltonian constraint $\hat H_q |\psi(Z_1,\dots,Z_J)\rangle=0$, which takes the same form as in (\ref{qH}) with $\hat q_j$ being a differential operator in $Z_i$, given by (\ref{qZ}).

This is the end of the quantization procedure. In the next section we demonstrate it by explicit examples for $J=1,2$.
In section \ref{sec5} we prove for general $J$ that this quantum theory is exactly equivalent to the four-dimensional fishnet theory in the ${\mathfrak u}(1)$ sector. 

\section{Tests and Examples} \la{sec4}
The aim of this section is to demonstrate by simple examples 
our quantization procedure. We will also be able to reproduce non-trivial exact results known from the planar fishnet CFT to all loop orders. In this section we restrict ourselves to the case $d=4$, $D=6$ corresponding to the initial problem.

\subsection{Single-particle case, with twist}\la{sec:J1}
In this section we consider the simplest example of $J=1$. 
If we literally set $J=1$ in the construction above we will get a contradiction, as in this case the Hamiltonian $H_q=-1$, due to the normal ordering. To make it non-trivial, but still simple, we deform the periodic conditions $X^M_{J+1}\equiv X^M_1$ by incorporating a twist
\beq
X_{J+1}=T.X_1
\eeq
where $T$ is a $6\times 6$ matrix
\beq\la{TMN}
{T^M}_N=
\left(
\begin{array}{cccccc}
	1 & 0 & 0 & 0 & 0 & 0 \\
	0 & 1 & 0 & 0 & 0 & 0 \\
	0 & 0 & \cos \theta  & -\sin \theta  & 0 & 0 \\
	0 & 0 & \sin \theta  & \cos \theta  & 0 & 0 \\
	0 & 0 & 0 & 0 & \cos \theta  & -\sin \theta  \\
	0 & 0 & 0 & 0 & \sin \theta  & \cos \theta  \\
\end{array}
\right)
\eeq
and $\theta$ is the twist parameter.
As we will see the twisted system has a non-trivial spectrum already for $J=1$. The fishnet CFT also needs to be twisted in the way explained in \cite{twistingpaper}. The Hamiltonian in this case is the same as before, with the only modification that the twist matrix appears under the trace
\beq
\hat H_q=\frac{1}{2}\tr(:q_1^2:T)-1\;.
\eeq
Note that since $:q_1^2:$ is a symmetric matrix only the symmetric part of $T$ contributes, which means that we can omit all the $\sin\theta$ entries in \eq{TMN}, retaining only the diagonal part. From that it is clear that the twist breaks the $SO(1,5)$ symmetry of the initial system down to ${\mathbb R}\otimes SO(4)$.
The convenient set of coordinates is thus
\beq\la{globalAdS}
Z^M=\{\cosh\rho\,\cosh s,\,\cosh\rho\,\sinh s,\,\sinh \rho\; \vec y\}
\eeq
where $\vec y$ is a unit 4D vector. The wave function is  a function of $\rho$, $s$ and $\vec y$. First, we notice that the dependence on $s$ is totally fixed by the ${\mathbb R}$ part of the symmetry
generated by $\hat q_1^{-1,0}=\frac{i}{\xi}\d_s$ with eigenvalue
$i\Delta/\xi$ determining the conformal dimension of the state $\Delta$. This implies the dependence on $s$ must be in the form of the factor $e^{\Delta s}$. 
Next, the wave function should belong to an irrep of $SO(4)$.
Having a single unit vector $\vec y$ we can only build a rank-$S$ symmetric traceless tensor out of it. It is convenient to introduce a null vector $\vec n=\{1,-i,0,0\}$, so that the highest weight states w.r.t. $SO(4)$ have the dependence on $\vec y$ of the form $(\vec y.\vec n)^S$. 
Thus, on symmetry grounds we deduce the following form of the wave-function
\beq\la{J1psi}
\psi(Z)=e^{s\Delta}(\vec y.\vec n)^S f(\rho)\;.
\eeq
Next we have to impose two constraints
\beq
\hat {\bf C}_2|\psi\> ={\bf C}_2|\psi\>\ ,\qquad \hat H_q|\psi\> = 0\;.
\eeq
Each of these constraints is a second-order differential operator.
For example
\beq\la{feq}
{1\over \psi}\,\hat {\bf C}_2\circ\psi=
{f''(\rho )\over f(\rho)}+{f'(\rho)\over f(\rho)} (\tanh \rho +3 \coth \rho )+ 
\frac{\Delta ^2}{\cosh^2\rho} -\frac{S (S+2)}{\sinh^2\rho} \;.
\eeq
The equation ${\bf C}_2+3=0$ has two solutions.
The solution which is regular at the origin takes the form
\beq
f_{\Delta,S}(\rho)=\frac{\tanh ^S\rho }{\cosh\rho} \,
   _2F_1\left(\frac{1}{2} (S-\Delta
   +1),\frac{1}{2} (S+\Delta +1);S+2;\tanh
   ^2\rho \right)\;.
\eeq
At small $\rho$ it behaves as $f_{\Delta,S}(\rho)\simeq \rho^S(1+{\cal O}(\rho^2))$.
The second solution can be obtained by the replacement $S\to -2-S+\epsilon$
and then taking the limit $\epsilon\to 0$, adjusting the normalization so that the limit is finite.
At large $\rho$ the regular solution $f_{\Delta,S}(\rho)$ have the following decaying asymptotic
\beq
f_{\Delta,S}(\rho)\simeq e^{-\rho }\frac{2  \Gamma (S+2)}{\Gamma
   \left(\frac{1}{2} (S-\Delta +3)\right)
   \Gamma \left(\frac{1}{2} (S+\Delta
   +3)\right)}+{\cal O}(e^{-3\rho})\;,
\eeq
which shows that this solution is well behaved at the boundary and decays as $1/Z^+$. We can form a combination of the two solutions which decays even faster as $1/(Z^+)^3$, however, it will diverge as $\rho^{-2-S}$ near the origin. We also found that there are $\log\rho$ terms in the small $\rho$ expansion, meaning that this  faster-decaying solution is not even single-valued.

Having the wave function completely fixed, we apply $\hat H_q$ to get
\beq\la{HqJ1}
\hat H_q\circ\psi = \(\frac{\sin ^2\left(\frac{\theta }{2}\right) (-\Delta +S+1) (\Delta +S+1)}{\xi ^2}-1\)\psi\;.
\eeq
In fact, to obtain \eq{HqJ1} we do not need to specify precisely which of the two solutions for $f(\rho)$ to use, since \eq{HqJ1} is also invariant under $S\to -2-S$. 
Finally, we see that \eq{HqJ1} fixes the spectrum as a function of the coupling $\xi$ and spin $S$
\beq\la{Dres}
\Delta = \sqrt{(S+1)^2-\frac{\xi ^2}{\sin ^2\frac{\theta }{2}}}\;.
\eeq
Notably, the weak coupling expansion of this expression has a meaningful form -- 
it starts from an integer bare dimension $S+1$ and goes in powers of $\xi^2$.
It is shown~\cite{twistingpaper} that \eq{Dres} is indeed the correct result for the spectrum of the {\it twisted} scalar in the fishnet CFT~\cite{twistingpaper}.

\subsection{Two-particle case}\la{J2sec}

To further explore how the quantum fishchain theory works, we consider the simplest example with no twist, namely $J=2$. 
In this case we have two particles with coordinates $Z_1$ and $Z_2$
and so the wave function is $\psi(Z_1,Z_2)$. 
According to our general procedure we have to impose for each particle one local constraint (for $d=4$)
\beq\la{J2local}
\hat{\bC}_{2,i}|\psi\rangle=-3|\psi\rangle\ ,\qquad i=1,2\;.
\eeq
In addition, we can choose the wave function to transform in an irreducible representation of the global conformal group, characterised in general by a conformal dimension $\Delta$ and a spin $S$. The highest-weight member of this multiplet then satisfies
\beq\la{J2global}
\hat{Q}^{-1,0}\,|\psi_{\Delta,S}\rangle=i\Delta\,|\psi_{\Delta,S}\rangle\ ,\qquad \hat{Q}^{1,2}\,|\psi_{\Delta,S}\rangle=S\,|\psi_{\Delta,S}\rangle\;.
\eeq
The highest-weight state by definition must be annihilated by the special conformal transformations $K^\mu$ and by the two raising operators of $SO(4)$ -- $\hat S^{+,a}$, which can be written as
\beq
\hat K^\mu={\cal X}_M \hat Q^{M,\mu}\ ,\qquad\mu=1,\dots,4\ ,\qquad
\hat S^{+,a}={\cal N}_M\hat  Q^{M,2+a}\ ,\quad a=1,2
\eeq
where ${\cal X}$ and ${\cal N}$ are two fixed null vectors
\beq\la{XNdef}
{\cal X}^M=\{\tfrac12,\tfrac12,0,0,0,0,0\}\ ,\qquad{\cal N}^M=\{0,0,1,-i,0,0\}\;.
\eeq
Thus the highest-weight wave function is a covariant combination of $Z_1$,  $Z_2$, ${\cal X}$ and ${\cal N}$. There are $5$ non-trivial structures one can build out of these $4$ vectors: $Z_1.Z_2,\;Z_a.{\cal X},\;Z_a.{\cal N}$, $a=1,2$. One can easily check that
$Z_1.Z_2,\;Z_a.{\cal X}$ are invariant under both $K^\mu$
and $S^{+,a}$, however, $Z_a.n$ transforms non-trivially under $\hat K^1$
and $\hat K^2$ and the invariant combination is $\frac{Z_1.{\cal N}}{Z_1.{\cal X}}-\frac{Z_2.{\cal N}}{Z_2.{\cal X}}$.
Next, we have to require \eq{J2global}, which fixes the dependence
of the wave function on $2$ out of remaining $4$ combinations:
\beq\la{psiansatz}
\psi_{\Delta,S}(Z_1,Z_2)=\frac{F\(Z_1.Z_2,\log\frac{Z_1.{\cal X}}{Z_2.{\cal X}}\)}{(Z_1.{\cal X}\;Z_2.{\cal X})^{\frac{\Delta-S}{2}}}\(\frac{Z_1.{\cal N}}{Z_1.{\cal X}}-\frac{Z_2.{\cal N}}{Z_2.{\cal X}}\)^S\;.
\eeq
The function of two variables $F$ can be fixed from two local constraints \eq{J2local}. The difference and the sum of the two constraints \eq{J2local} gives
\beqa\la{C2rel}
&&2 (\gamma -\cosh \kappa ) F^{(1,1)}+ (4-\Delta +S)
F^{(0,1)}-   (\Delta +S)\sinh \kappa\;F^{(1,0)}=0\\
\nn&&
2 \left(\gamma ^2-1\right) F^{(2,0)}+2 F^{(0,2)}+4
\sinh \kappa\;  F^{(1,1)}+2 F^{(1,0)} (\gamma 
(5-\Delta +S)+  (\Delta +S)\cosh \kappa)\\
\nn&&+\frac{1}{2} (2-\Delta +S) 
(6-\Delta +S) F=0\;
\eeqa
where $\gamma\equiv Z_1.Z_2$ and $\kappa\equiv \log \frac{Z_1.{\cal X}}{Z_2.{\cal X}}$. We were not able to find an explicit solution to this system. Luckily, we do not need to solve these equations. The action of the $H_q$ on $\psi_{\Delta,S}$ given by \eq{psiansatz} is
\beq\la{Hpsi2}
H_q|\psi_{\Delta,S}\rangle=\frac{1}{4}\,{\rm tr}\(:q_2^2::q_1^2:\)|\psi_{\Delta,S}\rangle-|\psi_{\Delta,S}\rangle=0\;.
\eeq
The above equation contains up to $4$ $q$'s meaning that we get a $4$th-order PDE on $F$.
The expression is rather long and is given in \eq{ddddF}. However, after 
using \eq{C2rel}, the condition \eq{Hpsi2} reduces to
\beq\la{spectrum}
\(\frac{(\Delta -S-2) (\Delta -S-4) (\Delta +S-2)
	(\Delta +S)}{16 \xi
	^4}-1\) F(\gamma ,\delta )=0\;,
\eeq
giving the exact spectrum in agreement with \cite{Grabner:2017pgm,Kazakov:2018qez}. 

Alternatively, we can solve (\ref{C2rel}) in the limit where both $Z_1$ and $Z_2$ approach the boundary of $AdS_5$. In this limit $\gamma\to\infty$ and $\kappa$ is kept fixed. The first equation in (\ref{C2rel}) tells us that $\kappa$ derivatives of $F$ scale as $1/\gamma$. Then, the second equation becomes
\beq\la{Feq}
\(\gamma\d_\gamma+{2+S-\Delta\over2}\)\(\gamma\d_\gamma+{6+S-\Delta\over2}\)F=0\;.
\eeq
It tells us that the two solutions behave as $\gamma^{\Delta-S-2\over2}$ and $\gamma^{\Delta-S-6\over2}$ at large $\gamma$. By plugging either of these asymptotic solutions into (\ref{Hpsi2}) and expanding it at large $\gamma$ we reconstruct (\ref{spectrum}). 

To pick the physical one we have to extend the solution into the bulk and demand that it stays regular. It turns out that the regular solution is the one that scale as $F\ \to\ (Z_1.Z_2)^{\Delta-S-2\over2}$ near the boundary. Correspondingly, the wave function asymptotes to the boundary the three point function between two scalar operators of dimension three and one spin $S$ operator of dimension $\Delta$
\beq\la{3pwf}
\psi_{\Delta,S}\simeq  \frac{1}{Z_1^+ Z_2^+ }{\(\cN.\cX_1\,\cX_2.\cX-\cN.\cX_2\,\cX_1.\cX\)^S\over(\cX_1.\cX)^{\Delta+S\over2}(\cX_2.\cX)^{\Delta+S\over2}(\cX_1.\cX_2)^{2-\Delta+S\over2}}
\eeq
where here $Z_i^M/Z^+_i\simeq\cX_i^M$ are the boundary points for large $Z^+_i$. On the r.h.s. we recognise the standard structure on the $3$-point correlator of two scalars and an operator with spin $S$ and conformal dimension $\Delta$. The fact that the regular solution to (\ref{C2rel}) decays as $\sim 1/Z_i^+$ near the boundary was demonstrated for $J=1$ in the previous section and will be proven for any $J$ in the next section.

In summary, we demonstrate in this section how the known exact spectrum of the fishnet model can be derived from the quantum fishchain for small $J$'s.  In the next section, we explore a construction which allows us to build the fishchain bulk wave function in terms of the CFT boundary wave function $\phi_{\cal O}$ \eq{wf}.

\section{General proof of the duality}\la{sec5}

The goal of this section is to demonstrate the equivalence between the quantum fishchain model and the planar 4D fishnet in the ${\mathfrak u}(1)$ sector for any length $J$ at the level of the single trace operators.
We build an explicit map between the wave function of the fishchain model $\psi_{\cal O}(Z_1,\dots,Z_J)$ and the 4D CFT wave function, given by the correlation function of $J$ scalars with a local operator $\varphi_{\cal O}(x_1,\dots,x_J)$ as defined in (\ref{wf}). 
The map which relates the two in a natural way is
\beq\la{themap}
\psi_{\cal O}(Z_1,\dots,Z_J)=\int \prod_{i=1}^J \frac{d^4 x_i}{-4\pi^2(Z_i.{\cal X}_i)^3}
\varphi_{\cal O}(x_1,\dots,x_J)
\eeq
where ${\cal X}^M_i= {\cal X}^M(\vec x_i)\equiv\(\frac{1+x_i^2}{2},\frac{1-x_i^2}{2},\vec x_i\)$. In the denominator of \eq{themap} one can recognise  the bulk-to-boundary propagator for a scalar particle of mass square $-3$, dual to the fishnet scalar. It satisfies the bulk equation of motion or equivalently, the constraint 
\beq
\(\hat{\bC}_{2,i}+3\)\frac{1}{(Z_i.{\cal X}_i)^3}=0\;,
\eeq
where $\hat{\bC}_{2,i}$ acts on $Z_i$. From that it follows that the  r.h.s. of \eq{themap} satisfies all local constraints \eq{J2local} by construction.
What remains to be shown is that $\hat H_q|\psi\rangle = 0$. 
The proof of this relation is the main focus of this section.

The main idea is to show that the action of $(\hat H_q+1)$
under the map \eq{themap} results in the action of the inverse of the graph-building operator $\hat B^{-1}$ on the CFT wave function $\varphi_\cO$. 
The inverse of the graph-building operator is defined as
\beq\la{Binvdef}
\hat B^{-1}=\prod_{i=1}^J(x_i-x_{i+1})^2\prod_{j=1}^J \frac{-\Box_j}{4\xi^2}\;.
\eeq
Due to (\ref{phys}), it acts in a simple way on the CFT wave function $\varphi_\cO$ 
\beq
\hat B^{-1} \varphi_{\cal O}(x_1,\dots,x_J)=\varphi_{\cal O}(x_1,\dots,x_J)\;,
\eeq
and implies that the fishchain wave function $\psi_{\cal O}$ is indeed annihilated by the Hamiltonian $\hat H_q$.

The key identity, which we prove below, is that 
\eq{Binvdef} can be equivalently written as
\beq\la{Bmain}
\hat B^{-1} = {\rm tr}\(
\prod_i\frac{:(\hat {\mathfrak q}_i)^2:}{2}\)
\eeq 
where ${\mathfrak q}_i^{NM}$ are the conformal generators acting on the 4D variable $x_i$ only, and the normal ordering is defined in exactly the same way as before in \eq{noq}, i.e.   
\beq
:(\hat {\mathfrak q}_i^2)^{NM}:
\;\equiv\;
(\hat {\mathfrak q}^2_j)^{NM}-\frac{2i}{\xi}\hat {\mathfrak q}_j^{NM}-\frac{1}{\xi^2}\eta^{NM}
\;.
\eeq
Then we show that the map \eq{themap}
transfers the action by 6D $q_i^{NM}$ on $Z_i$ into the action by ${\mathfrak q}_i^{NM}$ on $x_i$, ensuring that $\hat H_q|\psi\rangle=0$.

The derivation outlined above proves the equivalence of the two theories for the ${\mathfrak u}(1)$ sector. 

\subsection{Proof of the duality at the quantum level}
Here we present the details of the construction outlined above.
\paragraph{Graph building operator in Embedding space.}
In order to prove the key identity \eq{Bmain} we work with embedding coordinates $X$, (\ref{flat}). We are going to introduce some useful notations which would allow us to deal with the $SO(1,5)$ symmetry in covariant way. For that we will uplift some 4D integrals into 6D.\footnote{See \cite{SimmonsDuffin:2012uy} for a brief review of this formalism.}
We first demonstrate the main principle on 
a simple example. Consider the integral 
\beq\la{Ifini}
F(Z)\equiv \int d^4 x \frac{f(\vec x)}{-4\pi^2(Z.{\cal X})^3}=\int \(d^6 X \;\delta(X^2)\;\delta(X^+-e^\alpha)\)\;\frac{f(\vec X/X^+)}{-4\pi^2{(Z.X)^3}}
\eeq
where $\alpha\in{\mathbb R}$. To show that the above identity is true, first, we notice that for $\alpha=0$, $X$ becomes ${\cal X}(\vec x)=\(\frac{1+x^2}{2},\frac{1-x^2}{2},\vec x\)$ upon resolving the $\delta$-functions. Furthermore, if $\alpha\neq 0$, we can 
perform the change of the integration variables $X^M\to e^\alpha X^M$, which would give us back the initial integral with $\alpha=0$.
One would refer to $X^+=1$ as a gauge choice in the CFT literature, \cite{Aharony:1999ti}. Next, since $F(Z)$ does not actually depend on $\alpha$
we can integrate \eq{Ifini} in $\alpha$ over the range $[-\Lambda,\Lambda]$, which would remove the second $\delta$-function and produce an extra $1/X^+$ factor.
Dividing the result by $2\Lambda$ we should get back $F(Z)$. Note that the result should not depend on $\Lambda$, so we can send it to infinity to get
\beq\la{embeddingint}
F(Z)=\lim_{\Lambda\to\infty}{1\over2\Lambda}
\int\limits_{e^{-\Lambda}\leq X^+\leq\, e^\Lambda} d^6 X \;\delta(X^2)\;\frac{f(\vec X/X^+)}{-4\pi^2X^+{(Z.X)^3}}\equiv \int D^4 X\frac{f(\vec X/X^+)}{-4\pi^2X^+(Z.X)^3}\;.
\eeq
The advantage of this form of the integral is that the integration measure is explicitly invariant under ${\mathfrak so}(1,5)$. The integration limits could potentially spoil the invariance. However, at least in our case, the dependence on the integration domain is suppressed at large $\Lambda$, as we discuss below. 

Using these notations we get the following formula for the bulk wave function
\beq\la{bwf}
\psi_{\cal O}(Z_1,\dots,Z_J)=
\int \prod_{i=1}^J \frac{D^4 X_i}{-4\pi^2(Z_i.X_i)^3}
\Phi_\cO(X_1,\dots,X_J)\;,
\eeq
where
\beq
\Phi_\cO(X_1,\dots,X_J)\equiv\frac{\varphi_{\cal O}(\vec X_1/X_1^+,\dots,\vec X_J/X_J^+)}{X_1^+\dots X_J^+}\;.
\eeq

Next, in order to uplift the inverse of the graph-building operator we will also need to rewrite the 4D d'Alembertian in terms of $X$. In fact  it is easy to see that  we can
simply replace the 4D d'Alembertian by the 6D due to the identity
$
\Box^{(6)}=\Box^{(4)}-4\d_{X^+}\d_{X^-}
$
and the fact that  $f$ does not depend on $X^-$.\footnote{For comeliness, in (\ref{Greeninembedding}) we write the corresponding Green function in embedding space.}

Combining these observations together we get the following identity
\beq
\la{BG}\int \prod_{i=1}^J \frac{d^4 x_i}{(Z_i.{\cal X}_i)^3}
\hat B^{-1} \varphi_{\cal O}(\vec x_1,\dots,\vec x_J)=
\int \prod_{i=1}^J \frac{D^4 X_i}{(Z_i.X_i)^3}
\hat {\cal B}^{-1} \Phi_\cO(X_1,\dots,X_J)\;,
\eeq
where $\hat{\cal B}^{-1}$ is an analog of $\hat B^{-1}$ written in terms of $X_i$
\beq\la{inverse}
\hat\cB^{-1}\equiv\prod_i X_i.X_{i+1}\prod_j{1\over2\xi^2}\Box^{(6)}_{j}\;.
\eeq
As a next step we rewrite $\hat\cB^{-1}$ in terms of the charge density.
\paragraph{Expressing $\hat\cB^{-1}$ in terms of the charge density.}
In order to distinguish the charge density operator acting as a differential operator on $X_i$
from $\hat q_i$, which acts on $Z_i$ we introduce a new notation 
\beq
\hat \bq_i^{M\,N}\equiv  \hat X_i^N\hat K_i^M- \hat X_i^M\hat K_i^N\ ,\qquad
\hat K_i^M\equiv  -\frac{i}{\xi}\eta^{MN}\frac{\d}{\d X_i^N}\;.
\eeq
Here $\hat X_i$ acts  on the functions of $X_i$ in the standard way by multiplication by $X_i$. It should not be confused with $\hat X_i$ introduced in the quantization procedure section~\ref{sec:fcq}.
Next, we evaluate explicitly the following combination of $\bq$'s 
\beqa
\nn\(\hat \bq_i^2-\frac{2i}{\xi}\hat \bq_i\)^{M\,N}&=&-\hat X_i^M\hat X_i^N\,\hat K_i^2-\hat X_i^2\,\hat K_i^M\hat K_i^N+\hat X_i^M\hat K_i^N(\hat X_i.\hat K_i)+\hat X_i^N\hat K_i^M(\hat X_i.\hat K_i)\\
\la{qsquare}&&-{i\over\xi}\eta^{MN}(\hat X_i.\hat K_i)-{i\over\xi}\hat X_i^M\hat P_i^N
-{i\over\xi}\hat X_i^N\hat P_i^M\;.
\eeqa
Let us plug this combination of $\bq's$ instead of $\hat {\cal B}^{-1}$ under the integral in the r.h.s. of \eq{BG}.
The first simplification we see is that $\hat X_i^2$ can be set to zero due to the $\delta(X_i^2)$ in the integration measure.
Furthermore, the operator $i\xi(\hat X_i.\hat K_i)$ simply counts the powers of $X_i$, so acting on $\Phi_{\cal O}$ we can simply replace $(\hat X_i.\hat K_i)\to \frac{i}{\xi}$, as it  acts non-trivially on the  $1/X_i^+$ factor only. We conclude that under the integral we have most of the terms cancelling and the result simplifies to
 
\beqa
\nn\(\hat \bq_i^2-\frac{2i}{\xi}\hat \bq_i\)^{M\,N}&\simeq&-\hat X_i^M\hat X_i^N\,\hat K_i^2+{1\over\xi^2}\eta^{MN}\;.
\eeqa
From that, we see that under the integral
\beqa\la{Btoq}
{\rm tr}\prod_i\frac{:\hat \bq_{i}^2:}{2}&\simeq&\tr\prod_i\(-\tfrac{1}{2}\hat X_i^M\hat X_i^N\,\hat K_i^2\)=\prod_i\hat X_{i}.\hat X_{i-1}\prod_i \frac{\Box^{(6)}_i}{2\xi^2}=\hat {\cal B}^{-1}
\eeqa
where the normal ordering is defined the same way as before $:\hat\bq_i^2:\equiv \hat \bq_i^2-\frac{2i}{\xi}\hat \bq_i-\frac{1}{\xi^2}$.

\paragraph{Map to the fishchain Hamiltonian.}
It remains to relate $\hat B$ with the bulk Hamiltonian $\hat H_q$.
We are going to show that
\beq\la{finally}
\int \prod_{i=1}^J \frac{D^4 X_i}{-4\pi^2(Z_i.X_i)^3}
\hat {\cal B}^{-1}\, \Phi_\cO(X_1,\dots,X_J)=(\hat H_q+1)
\psi_{\cal O}(Z_1,\dots,Z_J)\;.
\eeq
The equality (\ref{finally}) immediately follows from (\ref{Btoq}) and the identity 
\beq
\int \prod_{i=1}^J \frac{D^4 X_i}{-4\pi^2(Z_i.X_i)^3}
\, \hat{\bf q}_i\, F(X_1,\dots,X_J)=
\hat q_i\,\int \prod_{i=1}^J \frac{D^4 X_i}{-4\pi^2(Z_i.X_i)^3}
\,  F(X_1,\dots,X_J)\;,
\eeq
where $\hat \bq_i$ and $\hat q_i$ are the conformal generators,
where the first acts on $X_i$ and the latter on $Z_i$.
This identity follows from the co-variance of the integration measure.
To formally derive this almost obvious identity one performs integration by parts and notes that the boundary terms are suppressed by an infinite volume factor $\Lambda$ in (\ref{embeddingint}).

This completes the proof of the duality between the fishnet CFT and the fishchain at the quantum level.

\paragraph{Wave function properties.}
Here we discuss some properties of the map between the bulk wave function $\psi_{\cal O}$ and
what we call the boundary CFT wave function $\varphi_{\cal O}$.

First,
we notice the following property of $\psi_{\cal O}(Z_1,\dots,Z_J)$. Imagine we send one $Z_k$ far from other points $Z_i$, then the main contribution to the integral \eq{themap} will be received from the domain where $x_k$ is far from other integration points. Since the 4D wave function $\Phi_{\cal O}(X_1,\dots,X_J)$ is defined as a correlator of local primary operators with a local operator ${\cal O}$, in this limit we should have $\Phi_{\cal O}\simeq \frac{1}{X_k.{\cal X}(\vec 0)}$. Then, using that
\beq
\int  \frac{D^4 X_k}{-4\pi^2(Z_k.X_k)^3(X_k.{\cal X}(\vec 0))}=\frac{1}{Z_k.{\cal X}(\vec 0)}
\eeq
we see that the dependence on $Z_k$ in $\psi_{\cal O}(Z_1,\dots,Z_J)$
in this limit
should come as a factor $\frac{1}{Z_k.{\cal X}(\vec 0)}$.

Another important limit is when $Z_i$ approached the boundary.
As we show in appendix~\ref{App:nearboundary} in the limit $Z_{i}^+\to\infty$
from \eq{bwf}
we get the following asymptotic
\beq\la{bulktoboundarylimit}
\psi_{\cal O}(Z_1,\dots, Z_J)\simeq\frac{\phi_{\cal O}(\vec Z_1/Z_1^+,\dots,\vec Z_J/Z_J^+)}{Z_1^+,\dots,Z_J^+}=\Phi_\cO(Z_1,\dots,Z_J)\;.
\eeq
We see that the bulk wave function approaches the boundary wave function with unit coefficient.
Thus for a generic point at the boundary, the wave function should decay as $1/Z_i^+$.
This is indeed the behaviour we found in section~\ref{sec:J1} for the case $J=1$ and in section~\ref{J2sec} for $J=2$.
The second solution of the local constraint $\hat{\bf C}_{2,i}=-3$, which decays faster at the boundary as $1/(Z_i^+)^3$, should have a singularity in the bulk as we demonstrated in~\ref{sec:J1}. Indeed the integral \eq{bwf} does not have any additional divergences when $Z$ gets deeper into the bulk. We conclude that regularity in the bulk is responsable for selecting the correct physical solution for $\psi_{\cal O}(Z_1,\dots, Z_J)$. 

\paragraph{The fishchain norm.}
We close this section by constructing the fishchian norm. As discussed in the introduction, the CFT norm in the ${\mathfrak u}(1)$ sector is given by (\ref{CFTnorm}), which we rewrite here in the new notations as 
\beq\la{boundarynorm}
\<\!\!\<\Phi_{\widetilde\cO}|\Phi_\cO\>\!\!\>=\int\prod_{i=1}^J {D^4 X_i\over4\pi^2}\;
\Phi_{\widetilde\cO}^\dagger(X_1,\dots,X_J)\prod_j\(-\Box_j\)\,\Phi_\cO(X_1,\dots,X_J)
\eeq

We now define the corresponding ``{\it fishchain norm}", $\<\!\<\psi_{\widetilde\cO}|\psi_\cO\>\!\>$ such that it is equal to $\<\!\!\<\Phi_{\widetilde\cO}|\Phi_\cO\>\!\!\>$. One can always use (\ref{bulktoboundarylimit}) to relate the bulk wave-function to the boundary one and then plug the results in (\ref{boundarynorm}). Here however, we construct a direct and more natural map that we refer to as the ``{\it fishchain norm}". 

Consider the regularized integral 
\beq
S[\psi_{\widetilde\cO},\psi_\cO;\{\Lambda_i\}_{i=1}^J]\equiv\int\limits_{Z_i^+<\Lambda_i} \prod_{i=1}^J{D^5 Z_i\over4\pi^2}\,
\psi_{\widetilde\cO}^\dagger(Z_1,\dots,Z_J)\prod_{j=1}^J\(\overleftarrow\nabla_{j\,M}\overrightarrow\nabla_j^M-3\)\psi_\cO(Z_1,\dots,Z_J)\ ,
\eeq
where $D^5 Z_i\equiv d^6 Z_i\; \delta(Z_i^2-1)$ is the integration measure of a unit radius AdS$_5$ space and
\beq\la{AdSderivative}
\overrightarrow\nabla_{j\ M}={\d\over\d Z^M_j}+Z_{j\,M}\(Z_j^N{\d\over\d{Z_j^N}}\)
\eeq
is the covariant derivative in AdS$_5$. This integral looks like the direct product of the on-shell actions for $J$ scalars in AdS$_5$ with mass $m^2=-3$. Upon integrating by parts the expression inside the brackets in \eq{bulknorm} becomes $-\nabla_{j\,M}^\dagger \nabla_{j}^M-3=-(\hat{\bf C}_{j,2}+3)=0$ so that only the boundary term survives. In appendix \ref{App:nearboundary} we show that at large $\Lambda_i$ that boundary term is given by\footnote{Here, we assumed that $\<\!\<\varphi_{\widetilde\cO}|\varphi_{\cO}\>\!\>$ is finite.}
\beq\la{boundaryS}
S=-\int\prod_{i=1}^Jd^4x_i\,\varphi_{\widetilde\cO}^\dagger(x_1,\dots,x_J)\prod_{j=1}^J\(\Lambda_j^2+\log(\Lambda_j^2)\,\Box_j+O(1)\)\varphi_\cO(x_1,\dots,x_J)\ .
\eeq
Hence, the fishchain norm is given by the coefficient of the  log-divergent piece for each scalar
\beq\la{bulknorm}
\<\!\<\psi_{\widetilde\cO}|\psi_\cO\>\!\>\equiv\(\prod_{j=1}^J\lim_{\Lambda_j\to\infty}{1\over4}\(2-\Lambda_j\d_{\Lambda_j}\)\Lambda_j\d_{\Lambda_j}\)S[\psi_{\widetilde\cO},\psi_\cO;\{\Lambda_i\}_{i=1}^J]\ .
\eeq
It is the only regularisation scheme-independent piece. Namely, we could equivalently use any other radial cutoff in AdS$_5$ and the coefficient of the logarithm in (\ref{boundaryS}) is not affected.

Finally, we note that $\xi^{2J}\,\hat\cB^{-1}$ in (\ref{inverse}) is self-adjoint with respect to the norm (\ref{boundarynorm}). Correspondingly, its eigenvalues are all real. The violation of unitarity comes about when equating them to 1 in (\ref{phys}) and solving for $\Delta$. Correspondingly, on the dual fishchain side, the operator $\xi^{2J}\,(\hat H_q+1)$ in (\ref{qH}) is self-adjoint with respect to the norm (\ref{bulknorm}). One heuristic interpretation of the Hamiltomian constraint $\hat H_q|\psi\rangle=0$ is the fishchain analog of the level-matching condition in string theory, ensuring the periodicity of the string in spacetime. The fact that this condition is responsible for the violation of unitarity goes hand in hand with the worldsheet interpretation of $\gamma$-deformation as inducing a twist on the sphere. In the fishnet limit that twist is taken to be complex and unitarity is lost.

\section{Integrability}\la{sec6}
In this section we demonstrate how the quantum integrability arises in the bulk theory. First, we describe the classical integrability based on \cite{Gromov:2019aku} and then show that it is preserved at the quantum level.

\subsection{Classical Integrability}
The key element of the construction are the $4\times 4$ matrices ${\mathbb L}_k(u)$ defined by
\beq
{\mathbb L}_k(u)=u\;{\mathbb I}_{4\times 4}-\frac{i}{2} q_k^{MN} \Sigma_{MN}
\eeq
where $\Sigma_{MN}$ are the $\sigma$-matrices in $6$D. They realise irrep ${\bf 4}$ of the conformal algebra ${\mathfrak so}(1,5)$ and satisfy\footnote{We use the representation given in Appendix \ref{appendix}.}
\beq\la{qmn}
[\Sigma_{MN},\Sigma_{KL}]=
-i\eta_{MK}\Sigma_{NL}
+i\eta_{NK}\Sigma_{ML}
+i\eta_{ML}\Sigma_{NK}
-i\eta_{NL}\Sigma_{MK}\;.
\eeq
At the same time the charge density $q_k^{MN}$ satisfy the ${\mathfrak so}(1,5)$ algebra under the Poisson and the Dirac brackets
\beq
\{q^{MN}_k,q^{KL}_k\}=
-\eta^{MK}q_k^{NL}
+\eta^{NK}q_k^{ML}
+\eta^{ML}q_k^{NK}
-\eta^{NL}q_k^{MK}\;.
\eeq
We note that in application to the current density there is no difference between the Poisson and Dirac brackets, as $q_i$ Poisson-commutes with all $2^{\rm nd}$ class constraints. Below we just use $\{\cdot,\cdot\}$ notation without making a distinction between these two brackets. 

As a consequence of \eq{qmn} it is easy to check that the Poisson bracket between the matrix elements of ${\mathbb L}_k(u)$ is
\beq\la{LL0}
\{{\mathbb L}^{ab}_n(u),{\mathbb L}^{cd}_m(v)\}=
\frac{{\mathbb L}^{cb}_n(u) {\mathbb L}^{ad}_n(v)-
	{\mathbb L}^{ad}_n(u) {\mathbb L}^{cb}_n(v)}{u-v}\delta_{nm}
\eeq
or, equivalently, in the integrability literature notations
\beq\la{LL}
\{{\mathbb L}_n(u)\;\overset{\otimes}{,}\;{\mathbb L}_m(v)\}=[r(u,v),{\mathbb L}_n(u)\otimes {\mathbb L}_n(v)]\delta_{nm}
\eeq
where the matrix $r(u,v)=\frac{{\mathbb P}}{u-v}$ and ${\mathbb P}$ is the $16\times 16$ permutation matrix. Acting on a tensor product of two 4D vectors ${\mathbb P}$ simply interchanges the vectors.

The main consequence of \eq{LL} is that the T-functions defined as
\beq
{\mathbb T}^{\bf 4}(u)={\rm tr}\[{\mathbb L}^{\bf 4}_J(u){\mathbb L}^{\bf 4}_{J-1}(u)\dots {\mathbb L}^{\bf 4}_1(u)\]
\eeq
are Poisson-commuting i.e. $\{{\mathbb T}^{\bf 4}(u),{\mathbb T}^{\bf 4}(v)\}_{PB}=0$ (see for example \cite{Faddeev:1996iy}). In order to demonstrate the integrability we should also show that the T-functions Poisson-commute with the Hamiltonian of the fishchain. To embed the Hamiltonian into the integrability construction, we have to follow the so-called {\it fusion} procedure (see for example \cite{Lipan:1997bs}). This construction allows one to build T-functions in irrep $\bf 6$ and $\bf \bar 4$. These additional T-functions, together with the ${\mathbb T}^{\bf 4}$ form the complete set of Poisson-commuting quantities.
In particular we show that ${\mathbb T}^{\bf 6}$ contains the Hamiltonian $H_q$ of the fish chain. 

To construct $T$-function in irrep $\bf 6$ we use simply that ${\bf 4}\otimes{\bf 4}
={\bf 6}\oplus {\bf 10}$
\beq
{\mathbb L}^{\bf 4}_k(u)\otimes
{\mathbb L}^{\bf 4}_k(u)={\mathbb L}^{\bf 6}_k(u)\oplus {\mathbb L}^{\bf 10}_k(u)\;.
\eeq
Next, we project into $\bf 6$ by antisymmetrizing w.r.t. the two spaces. Similarly one can construct $\bar {\bf 4}$ from the tensor product of $3$ copies of $\bf 4$.
For compliteness we also construct the L-matrix in the trivial (determinant) representation $\bar {\bf 1}$. This
 procedure results in
\beqa
\nn{\mathbb L}^{\bf 6}_k(u)&=&\(u^2-\tfrac{1}{8}\tr\! q_k^2\)+u q_k+\frac{q_k^2}{2}\;,\\
\la{clfu}{\mathbb L}^{\bf \bar 4}_k(u)&=&\(u^2-\tfrac{1}{8}\tr\! q_k^2\)\;\[-{\mathbb L}^{{\bf  4}}(-u)\]^T\ ,\\
\nn{\mathbb L}^{\bf \bar 1}_k(u)&=&\det{\mathbb L}^{{\bf  4}}(u)=\(u^2-\tfrac{1}{8}\tr\! q_k^2\)^2\ .
\eeqa
The above expression is general and does not assume any constraints. In our case, however, $X_k.X_k=P_k.X_k=0$ so that $\tr q_k^2=0$ and the above expressions can be simplified.
Using the result \eq{clfu} we can see that ${\mathbb T}^{\bf 6}$
contains the Hamiltonian at  zero value of the spectral parameter $u$
\beq\la{HTcl}
{\mathbb T}^{\bf 6}(0)={\rm tr}\[
\prod_i\frac{q_i^2}{2}
\]=H_q+1\;.
\eeq
This means that $H_q$ is a part of a large mutually Poisson-commuting family of operators.
From that we see that the model is indeed classically integrable.

Finally, we see from \eq{clfu} that ${\mathbb L}^{\bar{\bf 4}}$ and ${\mathbb L}^{{\bf 4}}$ are closely related. Let us derive the corresponding relation between ${\mathbb T}^{\bar{\bf 4}}(u)$ and ${\mathbb T}^{\bar{4}}(u)$. From \eq{clfu} we have
\beq
{\mathbb T}^{\bar{\bf 4}}(u)=\(-u^2+\frac{1}{8}\tr q_k^2\)^J{\rm tr}\[{\mathbb L}^{\bf 4}_1(-u){\mathbb L}^{\bf 4}_2(-u)\dots {\mathbb L}^{\bf 4}_J(-u)\]\ .
\eeq
We see that up to an explicit factor it coincides with ${\mathbb T}^{\bf 4}(-u)$ after reflecting the order of the particles. This implies that for the states which are symmetric under the reflection $i\to J-i+1$, those two T-functions are related by $u\to-u$ and the multiplication by $(-u^{2})^J$.

\subsection{Quantum Integrability}
Now we generalise the results of the previous section to the quantum level. The matrix elements of ${\mathbb L}^{\bf 4}_k$ become operators themselves, acting on the quantum Hilbert space, and no longer commute with each other
\beq
\hat {\mathbb L}_k^{\bf 4}(u)=u\,{\mathbb I}^{(\text{phys})}\otimes{\mathbb I}^{(\text{aux})}-\frac{i}{2}\hat q_k^{MN}\otimes \Sigma_{MN}\ .
\eeq
As a result instead of the relation \eq{LL0} we get its quantum analog
\beq\la{RTT}
R^{a\;b}_{a'b'}(u-v)\,
{{\hat{\mathbb L}}_{\;\;c}}^{b'}(v)
\hat{\mathbb L}^{a'}_{\;\;d}(u)
=
\hat{\mathbb L}_{\;\;d'}^{a}(u)\hat{\mathbb L}^{b}_{\;\;c'}(v)\,
R_{c\;d}^{c'd'}(u-v)
\eeq
also known as RLL-relation. 
For simplicity we ommited the site label $k$ and the representation label ${\bf 4}$.
The R-matrix is given by
\beq
R^{ab}_{cd}(u)=\delta^a_{c}\delta^b_{d}+\frac{i}{\xi}\frac{1}{u}
\delta^a_{d}\delta^b_{c}\;.
\eeq
The RLL-relation ensures that the T-operators at different values of the spectral parameter $u$ commute with each other as  operators on the Hilbert space. The T-operators are built in the same way as before i.e.
\beq\la{HT4}
\hat{\mathbb T}(u)=\tr\!\(
\hat{\mathbb L}_J(u)\,
\hat{\mathbb L}_{J-1}(u)\,\dots\,
\hat{\mathbb L}_1(u)\)\;.
\eeq
In order to check that our Hamiltonian $\hat H_q$ is a part of the commuting family of operators we have to build $\hat{\mathbb L}^{\bf 6}(u)$.
However, if we simply replace $q$ by $\hat q$ in the classical expression
\eq{clfu}, the corresponding T-operator will not commute with itself or with $\hat T^{\bf 4}(u)$. The right way of doing this
is to use the RLL relation \eq{RTT} as a building block, \cite{Lipan:1997bs}.  By noticing that $\tfrac12R(-i/\xi)$ is exactly an antisymmetrizer ${\mathbb P}$, which projects ${\bf 4}\times{\bf 4}$ onto ${\bf 6}$ we construct $\hat{\mathbb L}^{\bf 6}_k$
as $
{{\hat{\mathbb L}}_{\;\;c}}^{[b}\(u+\frac{i}{2\xi}\)
\hat{\mathbb L}^{a]}_{\;\;d}\(u-\frac{i}{2\xi}\)
$. One can check that this combination is also antisymmetric in $c,d$, which allows the consistent projection onto $\bf 6$, resulting in
\beq\la{L6q}
\hat{\mathbb L}^{\bf 6}_k=
\(u^2-\tfrac{1}{8}\tr\! \hat q_k^2\)+u \hat q_k+\(\frac{\hat q_k^2}{2}-\frac{i}{\xi}\hat q_k+\frac{1}{4\xi^2}\)\;.
\eeq
This expression generalizes the analogous classical expression \eq{clfu}. The classical limit can be recovered in the $\xi\to\infty$ limit, but in general \eq{L6q} contains some quantum corrections.
Again we can simplify the expression \eq{L6q} by taking into account the constraints specific for our model
$$\tr\! \hat q_k^2=-\frac{2}{\xi^2} \hat{\bf C}_{2,k}=\frac{6}{\xi^2}\;,$$
which allows to simplify it further to
\beq
\hat{\mathbb L}^{\bf 6}_k=
u^2+u \hat q_k+\frac12\({\hat q_k^2}-\frac{2i}{\xi}\hat q_k-\frac{1}{\xi^2}\)=u^2+u \hat q_k+\frac{:\hat q_k^2:}{2}\;.
\eeq
We can see now how the quantum Hamiltonian $\hat H_q$ appears in the integrability construction. Namely, we see that
\beq\la{HTqu}
\hat{\mathbb T}^{\bf 6}(0)=\tr\!\(
\hat{\mathbb L}^{\bf 6}_J(0)\,
\hat{\mathbb L}^{\bf 6}_{J-1}(0)\,\dots\,
\hat{\mathbb L}^{\bf 6}_1(0)\)={\rm tr}\[
\prod_i\frac{:\hat q_i^2:}{2}
\]=\hat H_q+1\ ,
\eeq
which shows that the our Hamiltonian $H_q$ is a part of the large mutually commuting family of operators which are thus integrals of motion. 

Finally, like in the classical case we can also construct $\hat{\mathbb L}^{\bf \bar 4}$ 
and $\hat{\mathbb L}^{\bf 1}$. The result is similar to the classical case but contains small extra shifts to the spectral parameters, which disappear in the classical limit~\cite{Lipan:1997bs}
\beqa
&&\(\hat{\mathbb L}^{\bf \bar 4}\)_a^{\;\;b}=\frac{\epsilon_{a\; a_1a_2a_3}\epsilon^{b\; b_1b_2b_3}}{3!}
{{\hat{\mathbb L}}_{\;\;\;b_1}}^{a_1}\(u+\frac{i}{\xi}\)
{{\hat{\mathbb L}}_{\;\;\;b_2}}^{a_2}\(u\)
\hat{\mathbb L}^{a_2}_{\;\;\;b_3}\(u-\frac{i}{\xi}\)\;,\\
&&\nn\hat{\mathbb L}^{\bf \bar 1}=\frac{\epsilon_{a_0 a_1a_2a_3}\epsilon^{b_0 b_1b_2b_3}}{4!}
{{\hat{\mathbb L}}_{\;\;\;b_0}}^{a_0}\!\!\(u+\frac{3i}{2\xi}\)
{{\hat{\mathbb L}}_{\;\;\;b_1}}^{a_1}\!\!\(u+\frac{i}{2\xi}\)
{{\hat{\mathbb L}}_{\;\;\;b_2}}^{a_2}\!\!\(u-\frac{i}{2\xi}\)
\hat{\mathbb L}^{a_2}_{\;\;\;b_3}\(u-\frac{3i}{2\xi}\)\;.
\eeqa
Evaluating these combinations explicitly we get:
\beqa
\hat{\mathbb L}^{\bf \bar 4}=\(u^2-\frac{\tr\!{\hat q_k^2}}{8}+\frac{1}{\xi^2}\)\[-\hat{\mathbb L}^{\bf  4}(-u)\]^T,\;\;
\hat{\mathbb L}^{\bf \bar 1}=\(u^2-\frac{\tr\!{\hat q_k^2}}{8}+\frac{5}{4\xi^2}\)^2+
\frac{\tr\!{\hat q_k^2}}{8\xi^2}-\frac{1}{\xi^4}
\eeqa
which simplify for $\tr\! \hat q_k^2=\frac{6}{\xi^2}$ it further reduces to
\beqa
\hat{\mathbb L}^{\bf \bar 4}=\(u^2+\frac{1}{4\xi^2}\)\[-\hat{\mathbb L}^{\bf  4}(-u)\]^T\ ,\qquad
\hat{\mathbb L}^{\bf \bar 1}=u^2\(u^2+\frac{1}{\xi^2}\)\;.
\eeqa
We see that like in the classical case, the T-operators for $\bar{\bf 4}$
and $\bf 4$ are related for the parity symmetric states by
$\hat {\mathbb T}^{\bar {\bf 4}}(u)=(-1)^J\(u^2+\frac{1}{4\xi^2}\)^{J}\hat {\mathbb T}^{\bar {\bf 4}}(-u)$. We also see that the {\it quantum determinant} for this model is $\hat {\mathbb T}^{\bar {\bf 1}}(u)=u^{2J}\(u^2+\frac{1}{\xi^2}\)^J$\footnote{The extra polynomial factors in T's can be removed by the gauge transformation $g(u)=e^{\frac{\pi  u}{2}} \Gamma (-i u)$. Under the gauge transformation
	$\hat {\mathbb T}^{\bar {\bf 1}}(u),\hat {\mathbb T}^{{\bf 1}}(u)\to u^J$, $\hat {\mathbb T}^{ {\bf 6}}(u)\to \hat {\mathbb T}^{ {\bf 6}}(u)/u^J$ and $\hat {\mathbb T}^{\bar {\bf 4}}(u)\to \hat {\mathbb T}^{\bar {\bf 4}}(u)/\(u^2+\frac{1}{4\xi^2}\)^J$ whereas ${\mathbb T}^{{\bf 4}}(u)$ does not change. The Q-operator transforms as ${\mathbb Q}(u)\to g(u) {\mathbb Q}(u)$.}.

We can perform the rough counting of the independent integrals of motion.
$\hat {\mathbb T}^{\bf 4}(u)$ is a polynomial of degree $u^J$ and thus contains $J-1$ non-trivial operatorial coefficients (the coefficient of $u^J$ is always four and $u^{J-1}$ is zero). Similarly $\hat {\mathbb T}^{\bf 6}(u)$ is an operator of degree $2J$ and contains $2J-1$ non-trivial operators.
In addition, $\hat {\mathbb T}^{\bar {\bf 4}}(u)$ provides further $J-1$ integrals so that in total we get order $\sim 4J$ independent integrals of motion, which matches the number of degrees of freedom of the fishchain.

We see that the quantum integrability of the fishchain model can be deduced rather easily in comparison to its CFT counterpart (see for comparison Appendix A of \cite{Gromov:2017cja}). This is due to the explicit $SO(1,5)$ covariance, which makes the derivation much more tractable. We expect that the separation of variables method, which is yet to be developed, can be applied in this dual formalism in a much more transparent way.

\section{Discussion}\la{sec7}

In \cite{Gromov:2019aku} we have derived the dual model to the fishnet CFT.
It is a weak to strong coupling duality between the single trace operators of the fishnet CFT in four dimensions and a quantum-mechanical system of particles forming a chain in five dimensions, to which we refer as fishchain. In \cite{Gromov:2019aku} we only considered the strong coupling limit, where the fishchain theory becomes classical. We have reproduced the strong coupling spectrum of short operators and even some non-trivial correlation functions, known exactly in the CFT. In this paper, we quantize the fishchain model and prove that the duality persists at the quantum level. We show that our result give the correct spectrum of short length operators at finite coupling, reproducing all loop order results from the CFT. Moreover, we have constructed an explicit map between the
wave function of the particles constituting the chain and some correlators in CFT valid in general for any operator length in the ${\mathfrak u}(1)$ sector. This map also directly relates the CFT graph building operator and the quantum Hamiltonian of the dual fishchain model, 
thus proving the duality at the quantum level. 

This is the first time
when a Holographic dual of a CFT in $D>2$ is derived from the first principles and the duality is proven at the quantum level in the planar limit\footnote{Our theory has no supersymmetries. Sometimes the supersymmetry helps to probe the holography exactly, but only for a  limited subset of BPS operators, preserving supersymmetries.}.

An important new feature we found in this paper is the emergence of the $AdS_5$ space at the quantum level. While the classical model was formulated on the lightcone of ${\mathbb R}^{1,5}$, we found that the quantum model naturally lives in $AdS_5$. The $AdS$ radius in string units appears to be quantum $R_{AdS}\sim \frac{1}{\xi}$. In the classical limit of  the fishchain, the `t Hooft coupling $\xi$ becomes large and reduces the AdS space to the lightcone.

In light of these results, it is important to emphasise the essence of planar holography and the lessons we learn from the explicit construction of a dual pair. 
What we found is much more than a change of variables. 
First, the holographic dictionary is such that the dual description becomes classical at strong coupling -- the inverse of the `t Hooft coupling $1/\xi$ plays the role of $\hbar$ in the fishchain model. It allows us to perform strong coupling computations by solving a classical system of coupled point particles in the lightcone of ${\mathbb R}^{1,5}$. Second, the holographic dual model should exhibit a certain degree of locality.
In the case of ${\cal N}=4$ SYM it is described in terms of a Polyakov action with a local interaction on the worldsheet. In our case, the fishchain is a sort of discretized string. Locality in such a discrete model exhibits itself in the canonical kinetic terms and in the nearest neighbours interaction.

There are many interesting and important future directions to pursue. We now discuss a few of them. 

First, for simplicity, in this paper, and in \cite{Gromov:2019aku} we have restricted our considerations to the so-called ${\mathfrak u}(1)$ subclass of operators. Namely, we considered operators that are only charged under one of the two $U(1)$ global internal symmetries. Extending the construction to any single trace operator in the model requires some extra work and will be reported in \cite{magnons}.

Second, the fishchain model we considered (\ref{fishnet}) is the simplest one in a large family of models \cite{Gurdogan:2015csr,Pittelli:2019ceq} for which {the holographic dual descriptions} are not yet known. A particularly interesting case is the supersymmetric versions of fishnets. It would be interesting to derive the fishchain duals of these models which are expected to include fermions and gauged supersymmetry in the bulk.

Third, it would be interesting to extend the duality beyond the strict planar limit by including systematic $1/N$ corrections. In the fishchain model, these corrections should be represented by a sort of fishchain vertex. Constructing such vertex is essential for a dual description of observable such as three-point functions, where the chain splits.\footnote{The fishchain vertex may also be a useful playground for understanding wrapping corrections in the Hexagon program for correlation functions, \cite{Basso:2015zoa,Basso:2018cvy}.}

Fourth, the fishnet model includes two types of single trace operators,
which we can call light and heavy. The heavy operators are those, whose conformal dimension is captured by the graph building operator and which grow at least as $\xi$ in the classical limit. Their dimension  $\Delta$ is a highly non-trivial function of $\xi$ and these are the operators we study in this paper. On top of these, there is an infinite tower of protected operators, whose dimension stays integer. They have zero anomalous dimension and their planar dynamics is free. Hence, in the planar limit, it is consistent to decouple them from the holographic description of the heavy dynamical single trace states. At higher order in the $1/N$ expansion, however, they are expected to start interacting. The dual description which goes beyond the strict planar limit must incorporate both of these types of states together with the fishchain vertex between them. It would be very interesting to understand the behaviour of these states beyond planar limit, as they also include the stress-energy tensor.

Fifth, the fishnet model can be viewed as a double scaling limit of $\gamma$-deformed ${\cal N}=4$ SYM theory, \cite{Gurdogan:2015csr,twistingpaper}. Now, that we have a first principles derivation of the dual for this theory, we can ask how to extend this derivation to the parent theory. To do so, we should turn back on the original string tension $\sqrt\lambda=R_{AdS}^2/l_s^2$. On the CFT side, this can be done systematically, order by order in $\lambda$. It would be interesting to work out these corrections for the fishnet and understand their holographic description, (see \cite{Correa:2012nk} for a cusped Wilson loops analogue and \cite{Bykov:2012sc} for a corresponding correction). 

Finally, analogous fishnet graphs also exist in other spacetime dimensions and it would be interesting to find their holographic duals. We can divide these types of graphs into two classes. The first class consists of graphs in which all propagators correspond to fields with canonical dimension. It includes, for example, the 3d and 6d graphs considered by Zamolodchikov \cite{Zamolodchikov:1980mb} related to the deformed ABJM model \cite{Caetano:2016ydc,Kazakov:2018gcy}. We expect the techniques used in this paper and in \cite{Gromov:2019aku} to generalize rather straightforwardly to this class of graphs. The second class requires non-local kinetic terms on the CFT side.  It includes, for example, the graphs of the SYK model \cite{Maldacena:2016hyu} as well as some of the generalized fishnet graphs considered in \cite{Kazakov:2018qez} and also BFKL limit of QCD. 
The non-locality on the boundary could require additional modification of our approach.
One sign of additional complexity of this case is that for this second class of graphs, the analog of the light states  do not decouple from the heavy states even in the strict planar limit. Correspondingly, we expect their dual description to be more involved but also, potentially, even more intriguing.

Another set of questions is related to the integrable structure of our model. In this paper we have shown that the fishchain model is quantum integrable. The model itself allows for a number of generalisations which would preserve integrability. First, one can introduce non-zero AdS radius already at the classical level and also consider a generalization of our model on a sphere.
One could hope that in a certain continuous limit $J\to\infty$
the fishchain would becomes smooth and could be described by a $\sigma$-model. By taking this limit we hope to reproduce the result of Zamolodchikov for the free energy~\cite{Zamolodchikov:1980mb} of the fishnet graphs, and reproduce the results of \cite{Basso:2018agi} obtained using the TBA methods. Importantly, with this discretization, one could be able to resolve the longstanding problem of quasntization of integrable $O(n)$ $\sigma$-models. Assuming this works one could then try to extend this approach to more complicated coset models which in particular describe strings in $AdS_5\times S^5$. So far all that one can compute in these models at the quantum level is the semiclassical spectrum around some classical solutions.\footnote{An alternative very promising approach to quantization of the string in a curved symmetric background is the pure spinors formalism~\cite{Berkovits:2019ulm}.} This direction could unlock the first principle derivation of the Zamolodchikov's S-matrix and formfactors in $\sigma$-models. It may also allow us to derive the Quantum Spectral Curve starting from the worldsheet theory.
Finally, the fishchain is the perfect playground for the development of the separation of variables method, which hopefully would allow us to solve the problem of computing of the correlation functions in planar ${\cal N}=4$ SYM theory.

\section*{Acknowledgements}

We thank V.~Kazakov, S.~Komatsu, L.~Rastelli, V.~Rosenhaus, K.~Zarembo and A.~Zhiboedov for invaluable discussions. We thank A.~Cavaglia, J.~Julius, and J.~Rostant for comments on the manuscript. N.G. was supported by the STFC grant (ST/P000258/1). A.S. was supported by the I-CORE Program of the Planning and Budgeting Committee, The Israel Science Foundation (1937/12) and by the Israel Science Foundation (grant number 968/15).

\appendix\la{appendix}
\section{Some more detailed equations}

This appendix contains some details that supplement the main text.

The equation \eq{Hpsi2} writen explicitly takes the form
{\small
	\beqa\la{ddddF}
	&&\frac{1 }{2 \xi ^4}F^{(0,4)}+\frac{2 \sinh \kappa  }{\xi ^4}F^{(1,3)}+
	\frac{(4 \gamma  \cosh \kappa+\cosh (2 \kappa )-5)  }{2 \xi ^4}F^{(2,2)}\\
	\nn &&+\frac{2 \left(\gamma ^2-1\right) \sinh \kappa  }{\xi ^4}F^{(3,1)}+
	\frac{\left(\gamma ^2-1\right)^2  }{2 \xi ^4}F^{(4,0)}\\
	%
	%
	\nn &&+\frac{(2 S+7) \cosh \kappa  }{\xi ^4}F^{(1,2)}+\frac{(\gamma  (3 S-\Delta +13)+(S+\Delta +1) \cosh \kappa) \sinh \kappa  }{\xi ^4}F^{(2,1)}\\
	\nn &&+\frac{\left(\gamma ^2-1\right) (\gamma  (S-\Delta +7)+(S+\Delta ) \cosh \kappa)  }{\xi ^4}F^{(3,0)}
	+\frac{(-S (S+2)-(\Delta -4) \Delta -8)  }{4 \xi ^4}F^{(0,2)}\\
	\nn &&+\frac{\left(S^2-2 (\Delta -8) S-\Delta ^2+24\right) \sinh \kappa  }{2 \xi ^4}F^{(1,1)}
	+\frac{(S-\Delta +2)^2 \left((S+\Delta )^2-4 (\Delta -3)\right) F }{32 \xi ^4}-F
	\\
	\nn &&+\frac{
		2  \left(\left(4 \gamma ^2+3\right) S^2
		+2 \left((34-6 \Delta ) \gamma ^2+5 \Delta -13\right) S
		+\Delta  (3 \Delta +10)+4 \gamma ^2 ((\Delta -16) \Delta +51)-64\right) F^{(2,0)} }{16 \xi ^4}\\
	\nn &&+\frac{ 
		4 \cosh\kappa \gamma  (2 S-2 \Delta +13) (S+\Delta )+
		\cosh({2 \kappa }) (S+\Delta +2) (S+\Delta )
	}{8 \xi ^4} F^{(2,0)}\\
	\nn &&+\frac{(-2 \gamma  (S-\Delta +5) (-4 S+(S+3) \Delta -9)-(S+\Delta ) (7 \Delta +S (2 \Delta -9)-24) \cosh \kappa) F^{(1,0)} }{4 \xi ^4}=0\;.
	\eeqa
}

One can verify that the familiar four-dimensional identity $\Box^{(4)}_x1/x^2=-4\pi^2\delta^4(x)$ becomes 
\beq\la{Greeninembedding}
\delta(X^2)\,\Box_X^{(6)}{1\over X.Y}=8\pi^2\int\limits_{{\mathbb R}^+}\! dc\,\delta^6(X-cY)
\eeq
in embedding space, which justifies the notation $\hat {\cal B}^{-1}$ in (\ref{inverse}). This is, however, not crucial for our derivation in section \ref{sec5}.

In section \ref{sec6} we have used the following representation for the $6D$ $\sigma$-matrices
\beqa
\begin{array}{ccc}
	
	\begin{array}{ccc}
		\Sigma _{1,2} & = & \left(
		\begin{array}{cccc}
			-\frac{i}{2} & 0 & 0 & 0 \\
			0 & -\frac{i}{2} & 0 & 0 \\
			0 & 0 & \frac{i}{2} & 0 \\
			0 & 0 & 0 & \frac{i}{2} \\
		\end{array}
		\right) \\
	\end{array}
	& 
	\begin{array}{ccc}
		\Sigma _{1,3} & = & \left(
		\begin{array}{cccc}
			0 & 0 & \frac{i}{2} & 0 \\
			0 & 0 & 0 & -\frac{i}{2} \\
			\frac{i}{2} & 0 & 0 & 0 \\
			0 & -\frac{i}{2} & 0 & 0 \\
		\end{array}
		\right) \\
	\end{array}
	& 
	\begin{array}{ccc}
		\Sigma _{1,4} & = & \left(
		\begin{array}{cccc}
			0 & 0 & -\frac{1}{2} & 0 \\
			0 & 0 & 0 & -\frac{1}{2} \\
			\frac{1}{2} & 0 & 0 & 0 \\
			0 & \frac{1}{2} & 0 & 0 \\
		\end{array}
		\right) \\
	\end{array}
	\\
	
	\begin{array}{ccc}
		\Sigma _{1,5} & = & \left(
		\begin{array}{cccc}
			0 & 0 & 0 & -\frac{i}{2} \\
			0 & 0 & -\frac{i}{2} & 0 \\
			0 & -\frac{i}{2} & 0 & 0 \\
			-\frac{i}{2} & 0 & 0 & 0 \\
		\end{array}
		\right) \\
	\end{array}
	& 
	\begin{array}{ccc}
		\Sigma _{1,6} & = & \left(
		\begin{array}{cccc}
			0 & 0 & 0 & -\frac{1}{2} \\
			0 & 0 & \frac{1}{2} & 0 \\
			0 & -\frac{1}{2} & 0 & 0 \\
			\frac{1}{2} & 0 & 0 & 0 \\
		\end{array}
		\right) \\
	\end{array}
	& 
	\begin{array}{ccc}
		\Sigma _{2,3} & = & \left(
		\begin{array}{cccc}
			0 & 0 & -\frac{i}{2} & 0 \\
			0 & 0 & 0 & \frac{i}{2} \\
			\frac{i}{2} & 0 & 0 & 0 \\
			0 & -\frac{i}{2} & 0 & 0 \\
		\end{array}
		\right) \\
	\end{array}
	\\
	
	\begin{array}{ccc}
		\Sigma _{2,4} & = & \left(
		\begin{array}{cccc}
			0 & 0 & \frac{1}{2} & 0 \\
			0 & 0 & 0 & \frac{1}{2} \\
			\frac{1}{2} & 0 & 0 & 0 \\
			0 & \frac{1}{2} & 0 & 0 \\
		\end{array}
		\right) \\
	\end{array}
	& 
	\begin{array}{ccc}
		\Sigma _{2,5} & = & \left(
		\begin{array}{cccc}
			0 & 0 & 0 & \frac{i}{2} \\
			0 & 0 & \frac{i}{2} & 0 \\
			0 & -\frac{i}{2} & 0 & 0 \\
			-\frac{i}{2} & 0 & 0 & 0 \\
		\end{array}
		\right) \\
	\end{array}
	& 
	\begin{array}{ccc}
		\Sigma _{2,6} & = & \left(
		\begin{array}{cccc}
			0 & 0 & 0 & \frac{1}{2} \\
			0 & 0 & -\frac{1}{2} & 0 \\
			0 & -\frac{1}{2} & 0 & 0 \\
			\frac{1}{2} & 0 & 0 & 0 \\
		\end{array}
		\right) \\
	\end{array}
	\\
	
	\begin{array}{ccc}
		\Sigma _{3,4} & = & \left(
		\begin{array}{cccc}
			\frac{1}{2} & 0 & 0 & 0 \\
			0 & -\frac{1}{2} & 0 & 0 \\
			0 & 0 & -\frac{1}{2} & 0 \\
			0 & 0 & 0 & \frac{1}{2} \\
		\end{array}
		\right) \\
	\end{array}
	& 
	\begin{array}{ccc}
		\Sigma _{3,5} & = & \left(
		\begin{array}{cccc}
			0 & -\frac{i}{2} & 0 & 0 \\
			\frac{i}{2} & 0 & 0 & 0 \\
			0 & 0 & 0 & -\frac{i}{2} \\
			0 & 0 & \frac{i}{2} & 0 \\
		\end{array}
		\right) \\
	\end{array}
	& 
	\begin{array}{ccc}
		\Sigma _{3,6} & = & \left(
		\begin{array}{cccc}
			0 & -\frac{1}{2} & 0 & 0 \\
			-\frac{1}{2} & 0 & 0 & 0 \\
			0 & 0 & 0 & -\frac{1}{2} \\
			0 & 0 & -\frac{1}{2} & 0 \\
		\end{array}
		\right) \\
	\end{array}
	\\
	
	\begin{array}{ccc}
		\Sigma _{4,5} & = & \left(
		\begin{array}{cccc}
			0 & \frac{1}{2} & 0 & 0 \\
			\frac{1}{2} & 0 & 0 & 0 \\
			0 & 0 & 0 & -\frac{1}{2} \\
			0 & 0 & -\frac{1}{2} & 0 \\
		\end{array}
		\right) \\
	\end{array}
	& 
	\begin{array}{ccc}
		\Sigma _{4,6} & = & \left(
		\begin{array}{cccc}
			0 & -\frac{i}{2} & 0 & 0 \\
			\frac{i}{2} & 0 & 0 & 0 \\
			0 & 0 & 0 & \frac{i}{2} \\
			0 & 0 & -\frac{i}{2} & 0 \\
		\end{array}
		\right) \\
	\end{array}
	& 
	\begin{array}{ccc}
		\Sigma _{5,6} & = & \left(
		\begin{array}{cccc}
			-\frac{1}{2} & 0 & 0 & 0 \\
			0 & \frac{1}{2} & 0 & 0 \\
			0 & 0 & -\frac{1}{2} & 0 \\
			0 & 0 & 0 & \frac{1}{2} \\
		\end{array}
		\right) \\
	\end{array}
	\\
\end{array}
\eeqa
and $\Sigma_{NM}=-\Sigma_{MN}$.

\section{Behaviour of the wave function near the boundary}\la{App:nearboundary}
Consider the function
\beq\la{IfiniA}
F(Z)\equiv \int d^4 x \frac{f(\vec x)}{(Z.{\cal X})^3}\ ,
\eeq
that resembles the holographic map (\ref{themap}). We use Poincar\'e coordinates to parametrize $Z=\{
Z^+\frac{1+z^2}{2}+\frac{1}{2Z^+},
Z^+\frac{1-z^2}{2}-\frac{1}{2Z^+},Z^+ \vec z\}$, so that $Z^2=-1$.
In the limit when $Z^+$ becomes large we approach the boundary of AdS. The following identity holds for general $Z^+$
\beq
 Z.{\cal X}=-\frac{1}{2}\((\vec z-\vec x)^2 Z^++\frac{1}{Z^+}\)\;.
\eeq
We see that the denominator in the integral \eq{IfiniA}
becomes very narrowly localised around $\vec x=\vec z$.
More precisely we have
\beq\la{deltad}
\frac{Z^+}{(Z.{\cal X})^3}=-4\pi^2 \delta^4(\vec x-\vec z)
-2\pi^2 \frac{\log Z^+}{(Z^+)^2}\,\Box_x\, \delta^4(\vec x-\vec z)+{\cal O}\(1/(Z^+)^{2}\)\ ,
\eeq
In particular from the leading term we find
\beq\la{themapA}
\psi_{\cal O}(Z_1,\dots,Z_J)= 
\frac{\phi_{\cal O}(z_1,\dots,z_J)}{Z_1^+\dots Z_J^+}+{\cal O}\(\frac{\log Z_i^+}{(Z_i^+)^3}\)
\equiv 
\Phi_{\cal O}(Z_1,\dots,Z_J)+{\cal O}\(\frac{\log Z_i^+}{(Z_i^+)^3}\)\;.
\eeq

\paragraph{Norm derivation.}
Consider the integral
\beq\la{bulknorm0}
I\equiv\int  d^6 Z\; \delta(Z^2+1)\;
\nabla^M F(Z)\,\nabla^MG(Z)\ ,
\eeq
where $\nabla_{M}$ is given in (\ref{AdSderivative}). Let us write this integral more explicitly in terms of the Poincar\'e coordinates $Z^+$ and $\vec z$ introduced above. In these coordinates we get
\beq
I=
\int d Z^+ d^4 \vec z \sqrt g\;
g^{ab}\d_a F(Z)\,\d_b G(Z)\ ,
\eeq
where the Poincar\'e metric $g_{ab}$ is a diagonal matrix ${\rm diag}\{1/(Z^+)^2,(Z^+)^2,\dots,(Z^+)^2\}$ so that $\sqrt g=(Z^+)^3$. Also $g^{ab}$ is the inverse metric, so that it is clear that the only term which is not suppressed at large $Z^+$ is the one with $a,b=+$:
\beq
I=
\int d Z^+ d^4 \vec z \;(Z^+)^5\;\d_+ F(Z)\,\d_+ G(Z)\ ,
\eeq
which gives the boundary term
\beq
\int d^4z \;(Z^+)^5\; F(Z)\,\d_+ G(Z)\ .
\eeq
Using that from \eq{deltad} 
\beq
F(Z)=\frac{f(\vec z)}{Z^+}+\log(Z^+)\frac{\Box f(\vec z)}{(Z^+)^3}+{\cal O}(1/(Z^+)^3)\ ,
\eeq
we get for the boundary term
\beq
I=-\int  d^4 \vec z\ 
(Z^+)^2\,f(\vec z)\, g(\vec z)
-2\log (Z^+)\int  d^4 \vec z \;\;
\,f(\vec z)\, \Box g(\vec z)+{\rm finite}\ .
\eeq
This relation is used to derive the fishchain norm, (\ref{bulknorm}).

\end{document}